\documentclass[10pt,twocolumn,letterpaper]{article}

\usepackage{cvpr}              

\usepackage[dvipsnames]{xcolor}

\usepackage{tabularx}

\definecolor{cvprblue}{rgb}{0.21,0.49,0.74}
\usepackage[pagebackref,breaklinks,colorlinks,citecolor=cvprblue]{hyperref}
\usepackage{makecell}
\usepackage[accsupp]{axessibility}
\usepackage{multirow}
\usepackage{float}

\def\paperID{***} 
\def\confName{CVPR}
\def\confYear{2024}

\title{OmniMedVQA: A New Large-Scale Comprehensive  \\ Evaluation Benchmark for Medical LVLM}

\author{Yutao Hu$^{1,2}$\footnotemark[1], \quad Tianbin Li$^{2}$\footnotemark[1], \quad Quanfeng Lu$^{2}$\footnotemark[1], \quad Wenqi Shao$^{2}$\footnotemark[2],\\ Junjun He$^{2}$, \quad Yu Qiao$^{2}$, \quad Ping Luo$^{1,2}$\footnotemark[2]\\
$^{1}$The University of Hong Kong \quad $^{2}$Shanghai AI Laboratory \\ 
}

\begin{document}
\maketitle

\footnotetext[1]{Equal contribution.}
\footnotetext[2]{Corresponding author.}

\begin{abstract}
Large Vision-Language Models (LVLMs) have demonstrated remarkable capabilities in various multimodal tasks. However, their potential in the medical domain remains largely unexplored. A significant challenge arises from the scarcity of diverse medical images spanning various modalities and anatomical regions, which is essential in real-world medical applications. To solve this problem, in this paper, we introduce OmniMedVQA, a novel comprehensive medical Visual Question Answering (VQA) benchmark. This benchmark is collected from 73 different medical datasets, including 12 different modalities and covering more than 20 distinct anatomical regions. Importantly, all images in this benchmark are sourced from authentic medical scenarios, ensuring alignment with the requirements of the medical field and suitability for evaluating LVLMs. Through our extensive experiments, we have found that existing LVLMs struggle to address these medical VQA problems effectively. Moreover, what surprises us is that medical-specialized LVLMs even exhibit inferior performance to those general-domain models, calling for a more versatile and robust LVLM in the biomedical field. The evaluation results not only reveal the current limitations of LVLM in understanding real medical images but also highlight our dataset's significance.  Our code with dataset are available at  \url{https://github.com/OpenGVLab/Multi-Modality-Arena}.
\end{abstract}    
\section{Introduction}
\label{sec:intro}

Recently, Large Vision-Language Models (LVLMs) have exhibited remarkable advancements across various domains, including embodied AI \cite{mu2024embodiedgpt}, autonomous driving \cite{xu2023drivegpt4}, and remote sensing \cite{kuckreja2023geochat}. Encouraged by their achievements, a growing number of LVLMs tailored for medical applications have emerged, claiming impressive performance across a wide spectrum of medical challenges \cite{moor2023med, wu2023towards, zhang2023pmc, li2023llava}. However, despite the growing attention these models have obtained, there has been a noticeable lack of comprehensive evaluations, particularly when it comes to real medical images, which strongly hinders a thorough understanding of their applicability and performance in medical contexts \cite{wu2023can}.

We attribute this challenge to the absence of a comprehensive and diverse evaluation benchmark, one that encompasses images captured from various modalities and covers a broad spectrum of human anatomies. In more detail, the ability to answer questions based on a given image is fundamental, yet critically important in evaluating the performance of LVLMs. To facilitate this purpose, a comprehensive Visual Question Answering (VQA) dataset is indispensable. However, as indicated in Table~\ref{tab:stacom}, the majority of existing VQA datasets suffer from size limitations. Moreover, many of them provide only a limited number of modalities and focus exclusively on specific aspects of human anatomy. Consequently, these datasets do not meet the requirements for a comprehensive evaluation of LVLMs in the medical domain.

\begin{table}[tbp]
  \centering
  \caption{The comparison of the number of modalities, images and question-answering items in different medical VQA datasets. $^\dag$ indicates we calculate the numbers by ourselves, without the official statistic could be directly adopted.}
  \vspace{-2mm}
    \begin{tabular}{l|c|c|c}
    \toprule[1pt]
    Dataset & \# Modalities  & \# Images & \# QA Items  \\
    \hline
    VQA-RAD \cite{lau2018dataset}   &  3 &  315 &  3515   \\
    SLAKE \cite{liu2021slake} &  3 &  642 &  14,028  \\
    Path-VQA \cite{xuehai2020pathvqa}   & 2$^\dag$ & 4998 &  32,799 \\
    VQA-Med \cite{ben2021overview}  & 5$^\dag$ & 4500 &  5500 \\
    \hline
    OmniMedVQA & 12 & 118,010 &  127,995 \\
    \bottomrule[1pt]
    \end{tabular}%
  \label{tab:stacom}%
  \vspace{-4mm}
\end{table}%

\begin{table*}
\centering
\small
\caption{\textbf{Comparison of Different LVLMs.} VE, ToP and TuP indicate the visual encoder, number of total parameters and tuning parameters, respectively. $^\dag$ indicates that the model is frozen. CC$^*$ consists of COCO \cite{chen2015microsoft}, CC3M \cite{sharma2018conceptual}, and CC12M \cite{changpinyo2021conceptual}.  CC, VG, SBU CY, and L400 indicate Conceptual Caption \cite{sharma2018conceptual,changpinyo2021conceptual}, Visual Genome~\cite{krishna2017visual}, COYO-700M \cite{kakaobrain2022coyo-700m} and LAION 400M~\cite{schuhmann2021laion}, respectively. LLaVA-I and G4L represent 158K multimodal instruction-following data in LLaVA \cite{liu2023visual} and data generated by GPT-4 for building an instruction-following LLMs \cite{peng2023instruction}. QA$^*$ denotes 13 question-answering datasets in InstructBLIP \cite{dai2023instructblip}.}
\label{tab:comparison-LVLMs}
\resizebox{\textwidth}{!}{%
\begin{tabular}{l|cccc|cc|cc}
\toprule
\multirow{2}{*}{Model} & \multicolumn{4}{c|}{Model Configuration} & \multicolumn{2}{c|}{Image-Text Data} & \multicolumn{2}{c}{Visual Instruction Data}  \\
\cmidrule{2-9}
& VE      & LLM   &ToP &TuP   & Source  & Size     & Source                  & Size      \\
\cmidrule{1-9}
BLIP2 \cite{li2023blip} & ViT-g/14$^\dag$      & FlanT5-XL$^\dag$  &4B &107M    & CC$^*$-VG-SBU-L400  & 129M     & -                  & - \\
LLaVA \cite{liu2023visual} & ViT-L/14$^\dag$     & Vicuna   &7B &7B   & CC3M  & 595K     & LLaVA-I                  & 158K \\
LLaMA\_Adapter\_v2 \cite{gao2023llama} & ViT-L/14$^\dag$      & LLaMA$^\dag$    &7B &63.1M    & L400  & 200M     & LLaVA-I+G4L                 & 210K \\
MiniGPT-4 \cite{zhu2023minigpt} & BLIP2-VE$^\dag$   & Vicuna$^\dag$    &7B & 3.1M     & CC-SBU-L400  & 5M     & CC+ChatGPT                 & 3.5K \\
mPLUG-Owl \cite{ye2023mplug} &ViT-L/14 & LLaMA$^\dag$  &7B & 1.1B    & CC$^*$-CY-L400  & 204M    & LLaVA-I                 & 158K \\ 
Otter \cite{li2023otter} &ViT-L/14$^\dag$ & LLaMA$^\dag$  &9B & 1.3B    & -  & -    & LLaVA-I                & 158K \\ 
InstructBLIP \cite{dai2023instructblip} &ViT-g/14$^\dag$ & Vicuna$^\dag$  &7B & 107M    & -  & -    & QA$^*$                & 16M \\ 
VPGTrans \cite{zhang2023transfer} &ViT-g/14$^\dag$ & Vicuna$^\dag$  &7B & 107M   & COCO-VG-SBU  & 13.8M    &CC+ChatGPT                 & 3.5K \\ 
\hline
Med-Flamingo \cite{moor2023med} & ViT-L/14$^\dag$ & LLaMA $^\dag$  & 8.3B & 1.3B  & MTB, PMC-OA  & 2.1M    & -  & - \\ 
RadFM \cite{wu2023towards} & ViT-3D & LLaMA  & 14B & 14B   & MedMD  & 16M    & RadMD                & 3M \\ 
MedVInT\_TE \cite{zhang2023pmc} & ResNet-50 & LLaMA$^\dag$  & 7B & 156.4M    & PMC-OA  & 1.64M    &PMC-VQA                 & 152k \\ 
LLaVA-Med \cite{li2023llava} & ViT-L/14$^\dag$ & Vicuna  & 7B & 7B  & PMC-15M  & 600K    & PMC-15M + GPT4                & 60K \\ 
\bottomrule
\end{tabular}%
}
\end{table*}

\begin{figure*}[tbp]
\centering 
\includegraphics[width=1\linewidth]{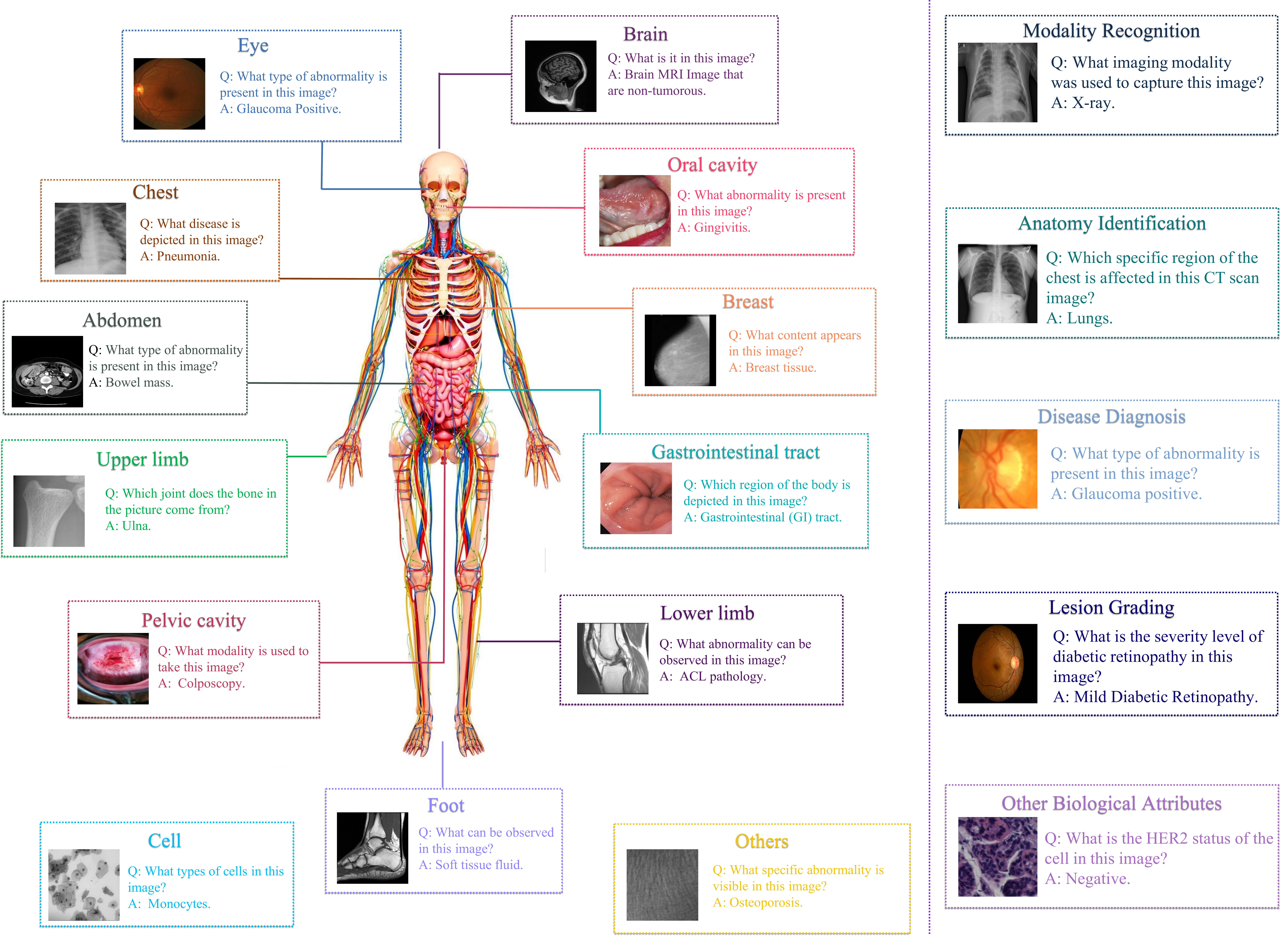} 
\caption{\textbf{Left:} Overview of our OmniMedVQA dataset. OmniMedVQA covers the majority of radiologic modalities and anatomical regions of the human body, such as the brain, eyes, oral cavity, chest, breast, abdomen, upper limb, lower limb, feet, etc. \textbf{Right:} Illustrations of samples from five different question types.}
\label{fig:body} 
\end{figure*}

To address this challenge, this paper introduces OmniMedVQA, a large-scale and comprehensive Visual Question Answering benchmark designed for the medical domain. Considering the scarcity of medical image-text data, we collect numerous medical classification datasets and then transfer these data to VQA format according to their classification attribute based on the powerful context reasoning capacity of GPT. Generally speaking, OmniMedVQA boasts two primary highlights. First, it encompasses images from 12 different modalities, including MRI, CT, X-Ray, histopathology, fundus photography, \emph{et al.}, resulting in a highly diverse dataset. Importantly, all these images originate from real medical scenarios, aligning OmniMedVQA closely with real-world applications. Second, OmniMedVQA covers over 20 distinct human anatomical regions. As illustrated in Fig~\ref{fig:body}, OmniMedVQA spans from the brain to the extremities, which facilitates a more comprehensive evaluation of different LVLMs and calls for a more versatile medical LVLM. Moreover, for the convenience of evaluation, we assign the incorrect options to each question-answering (QA) pair, transferring our OmniMedVQA to a multi-choice Question-Answer dataset. Overall, our OmniMedVQA dataset contains 118,010 different images with 127,995 different test items, leading to a large-scale evaluation benchmark.

In our evaluation, we assess a total of twelve representative models, including eight general-domain LVLMs, \emph{e.g.,} BILP2 \cite{li2023blip}, MiniGPT-4 \cite{zhu2023minigpt}, InstructBLIP \cite{dai2023instructblip}, mPLUG-Owl \cite{ye2023mplug}, Otter \cite{li2023otter}, LLaVA \cite{liu2023visual}, LLama\_adapter\_v2 \cite{gao2023llama}, and VPGTrans \cite{zhang2023transfer}, as well as four specialized medical LVLMs, including Med-Flamingo \cite{moor2023med}, RadFM \cite{wu2023towards}, MedVInT \cite{zhang2023pmc}, and LLaVA-Med \cite{li2023llava}. Notably, since OmniMedVQA is extremely challenging, especially for the general-domain LVLM, we find it is difficult for the model to directly generate the answer even if we give them the candidate options. To better evaluate their inherent knowledge in the biomedical domain, we adopt two different metrics, Question-answering score and Prefix-based Score \cite{xu2023lvlm, li2023blip} to calculate the VQA accuracy, leading to a more comprehensive evaluation. Through our extensive experiments, we surprisingly find, that medical-specialized LVLMs exhibit superior performance compared to general-domain LVLMs. Specifically, although medical LVLMs obtain better performance on some specific modalities such as CT, MRI and X-Ray, they struggle to consistently outperform general models across all modalities, particularly those with similar distributions to general images. Furthermore, we emphasize the pressing need for a robust model that can effectively align image-text pairs in the medical field. Such a model is crucial for medical-domain LVLMs, as it can generate accurate and comprehensive descriptions for medical images and support the sufficient training of LVLMs.

We want to emphasize that, although the classification attribute is compact, it does provide a basic evaluation benchmark for the medical area and mitigate the collecting cost. More importantly, according to the evaluation results, although not complex, the existing LVLMs, especially those medical-specialized LVLMs, do not exhibit satisfactory performance on our OmniMedVQA dataset, which not only shows the shortcoming of these models but also demonstrates the challenge of the proposed dataset.

The main contributions of this paper are summarized as follows:
\begin{itemize}	
\item We propose OmniMedVQA, a large-scale and comprehensive Visual Question Answering benchmark tailored to the medical domain. OmniMedVQA contains 12 different modalities and covers more than 20 unique human anatomical regions, establishing a comprehensive benchmark for evaluating the fundamental capabilities of LVLMs in addressing medical challenges. 
	
\item We conduct a thorough evaluation for 12 different LVLMs, including 8 general-domain LVLMs and 4 specialized LVLMs designed for medical applications. As far as we know, it is currently the most comprehensive evaluation of LVLMs towards the medical domain.

\item Our evaluation uncovers several innovative insights and provides valuable guidance for improving LVLMs toward medical applications in the future.

\end{itemize}
\section{Related Work}
\label{sec:related}

\subsection{Large Vision-Language Models}

Based on the recent emergence of Large Language Models (LLMs) such as LLaMA \cite{touvron2023llama} and GPT \cite{ouyang2022training}, LVLMs utilize the knowledge from LLMs and align visual features to the textual space for various text output. Flamingo \cite{alayrac2022flamingo} is one of the early attempts that insert cross-attention layers into LLMs to introduce visual features into textual space. Meanwhile, to better align multi-modal features, BLIP2 \cite{li2023blip} unifies the pre-trained visual encoder with LLM through an ingeniously designed Q-former. After that, InstructBLIP \cite{dai2023instructblip} extends BLIP-2 with instruction-following data and obtains better performance. Motivated by this success, most LVLMs are built through the instruction-tuning pipeline. For example, LLaVA \cite{liu2023visual} constructs 158K instruction-following data to conduct the training process and achieves great performance. Building upon the success of LLaVA, several subsequent LVLMs \cite{gao2023llama, li2023otter, ye2023mplug} leverage the high-quality 158k multimodal data to facilitate the training process. Furthermore, MiniGPT-4 \cite{zhu2023minigpt} aligns a frozen visual encoder with a frozen LLM via only one projection layer. To better fine-tune the model, MiniGPT-4 utilizes 3500 detailed image-description pairs, illustrating that even a relatively small amount of high-quality data can significantly enhance the training of LVLMs. Additionally, VPGTrans \cite{zhang2023transfer} transfers the text encoder of BLIP2 model to Vicuna, which reduces the training costs and maintains the convincing performance.

Recently, encouraged by the success of these general-domain LVLMs, researchers have embarked on the development of LVLMs for the medical field. Med-Flamingo \cite{moor2023med} is the pioneering effort in this field, which extends the Flamingo into the medical domain by pre-training on multi-modal knowledge sources across medical disciplines. Meanwhile, LLaVA-Med \cite{li2023llava} filter image-text pairs from PMC-15M \cite{zhang2023large}, and train a biomedical-specialized LVLM with a small amount data based on the LLaVA-pretrained parameter. Zhang \emph{et al.} generate a large-scale medical VQA dataset, PMC-VQA \cite{zhang2023pmc}, through the self-instruction on PMC-OA \cite{lin2023pmc}. Leveraging PMC-VQA, Zhang \emph{et al.} train a biomedical-specialized VQA model, termed MedVInT, which achieves state-of-the-art performance on many Medical VQA datasets. Furthermore, they continue to propose RadFM \cite{wu2023towards}, the first multi-modal foundation model for seamlessly integrating natural languages with both 2D and 3D radiologic images, which better fits the medical practical. Overall, we compare the information of aforementioned LVLMs in Table~\ref{tab:comparison-LVLMs}.

Despite the increasing attention, it still lacks a comprehensive evaluation of LVLMs within the medical domain. To address this deficiency, we perform a thorough evaluation for these LVLMs in this paper.

\subsection{Medical VQA Dataset}
With the rapid development of LVLMs, the field of Medical VQA has received considerable interest in recent years. VQA-RAD \cite{lau2018dataset}, SLAKE \cite{liu2021slake}, Path-VQA \cite{xuehai2020pathvqa}, and VQA-Med \cite{ben2021overview} are four widely used Medical VQA datasets. However, they all have less than 5K images. Meanwhile, VQA-RAD and SLAKE only contain images captured by CT, MRI or X-Ray, reducing their diversity and restricting their further application. Moreover, all these datasets only cover limited parts of the human anatomy, hindering a comprehensive evaluation across various human anatomical regions. For example, VQA-RAD only contains images of the head, chest, and abdomen, while Slake only covers the head, chest, abdomen, pelvis, and neck, lacking images of other important parts of the human body. Recently, motivated by the success of self-instruction in textual data generating, PMC-VQA, a large-scale Medical VQA dataset, is proposed based on numerous image-caption pairs. However, its images and text are extracted from online papers, which can result in image compression and a significant gap from real-world medical applications.

In this paper, we collect a new large-scale OmniMedVQA dataset, which contains medical images captured through 12 different modalities and covers almost every anatomical region of the human body. OmniMedVQA is the current largest MedVQA dataset with real medical images. We hope it can help the community better evaluate the fundamental ability of LVLMs in the biomedical field.

\section{Dataset Collection}
\label{sec:data}
In this part, we introduce the collection process of our OmniMedVQA dataset. To make full use of real medical images, we collect an enormous medical classification dataset and construct the question-answer pairs based on their inherent attributes using the ChatGPT API. Generally speaking, the construction process has the following four steps.

\begin{figure}[tbp]
\centering 
\includegraphics[width=1\linewidth]{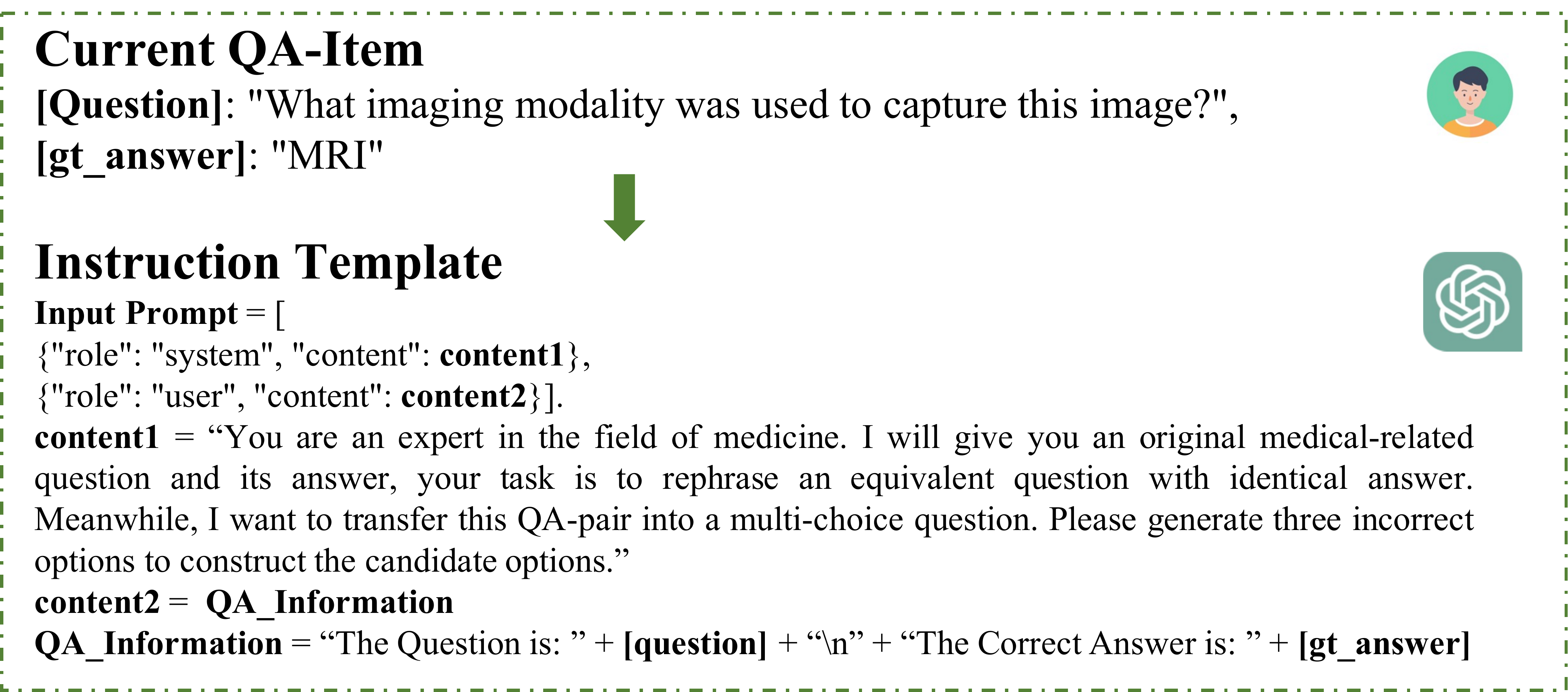} 
\caption{The illustration of the process by which we transfer the QA item from the QA template into the multi-choice question-answer pair.}
\label{fig:prompt} 
\end{figure}

\begin{itemize}	
\item \textbf{Original dataset preparing.} Due to the well-known difficulties in downloading medical datasets, it is notably time-consuming to obtain plentiful and suitable datasets. To construct a comprehensive VQA benchmark, we collected 73 different medical classification datasets encompassing 12 different imaging modalities, which span more than 20 different human anatomical regions. The details of all the involved datasets are presented in the supplementary material.

\item \textbf{Design QA templates.} Based on the collected datasets, we need to transfer the original classification attributes into QA format. To achieve this goal, we first construct the QA template for each dataset. On the one hand, the category information is naturally suitable for constructing a QA-pair. Therefore, we design question templates according to their original categories. For example, in the MAlig\_Lymph dataset\cite{orlov2010automatic}, it contains 3 different diseases. Therefore, we could design a QA template as ``Q: What is the specific diagnosis for the cancer cells in this image?; A: Chronic Lymphocytic Leukemia.''. On the other hand, through further understanding of the dataset, we construct QA pairs according to their other attributes, such as modality and anatomy information.  For example, in SARS-COV-2 Ct-Scan dataset\cite{soares2020sars}, we could also ask ``What is the modality of the image?'' or ``What is the abnormal organ in the picture?'', which evaluate the ability of modality recognition and anatomy localization. In summary, all QA pairs fall into five distinct question types: Modality Recognition, Anatomy Identification, Disease Diagnosis, Lesion grading and Other Biological Attributes. We illustrate the sample of each question type in Fig~\ref{fig:body}. As depicted in the right part of Fig~\ref{fig:body}, each question type could evaluate one specific capability within the biomedical field. Specifically, Lesion Grading aims to assess the severity of lesions in the images, while Other Biological Attributes include the analysis of various attributes related to medical images, such as cell shape, cancer status, imaging direction \emph{et al.} The detailed statistic information of each type is listed in Table~\ref{tab:statis}. Furthermore, we meticulously control the number of different items from each template to ensure balance and prevent significant bias. Specifically, we construct QA templates from 73 medical classification datasets and calculate the number of potential images that could be used under each template. Then, we select the images using the Inverse Proportional Sampling strategy. Namely, templates with a larger number of associated images are assigned a smaller sample ratio. In this way, our dataset keeps a balanced distribution across categories and avoids the bias on some reduplicate QAs, ensuring OmniMedVQA as a diverse and comprehensive dataset.

\begin{table}[tbp]
  \centering
  \caption{The number of images and items for different question types in OmniMedVQA.}
  \vspace{-2mm}
    \begin{tabular}{l|c|c}
    \toprule[1pt]
    Question Type & \# Images  & \# Items  \\
    \hline
    Modality Recognition   &  19,381 &  19,427   \\
    Anatomy Identification  &  19,992 &  20,330   \\
    Disease Diagnosis   & 73,099 & 73,455  \\
    Lesion Grading   & 2621 & 2621 \\
    Other Biological Attributes  & 12,156 & 12,162  \\
    \hline
    Total & 118,010 & 127,995 \\
    \bottomrule[1pt]
    \end{tabular}%
  \label{tab:statis}%
  \vspace{-4mm}
\end{table}%

\item \textbf{Refine QA pairs.}
In order to increase the diversity of our dataset and better evaluate the capability of each LVLM, we employ the ChatGPT-3.5 API to perform two key operations. First, we reformulate the question in each item to change the expression style and syntactic structure, while preserving the original semantic meaning, which allows us to evaluate the adaptability of LVLMs to various linguistic representations. Second, we leverage the GPT-3.5 API to generate a set of incorrect options for each item, which are utilized to construct multiple-choice question-answer pairs. Specifically, each item in our dataset is paired with incorrect answers as the candidate options. The number of options varies from 2 to 4, which depends on the content of the specific question. By doing so, it is more convenient to judge the correctness of the response from LVLM. The generation process of each item are depicted in Fig.~\ref{fig:prompt}.

\item \textbf{Human double check.} To ensure data quality, we conducted further inspections to guarantee the validity of our OmniMedVQA dataset.

\end{itemize}

Overall, OmniMedVQA contains 118,010 images with 127,995 QA-items, covering 12 different modalities and referring to more than 20 human anatomical regions. The detailed modality and anatomy information of our dataset are listed in Table~\ref{tab:info}. We want to emphasize that while the classification attributes may appear intuitive and not overly complex, they play a crucial role in evaluating the fundamental capabilities of LVLMs in the medical domain, which are critical to support broader applications in this field. Meanwhile, the evaluation results reveal that the medical-specialized LVLM cannot handle these questions well, indicating a deficiency in its foundational medical knowledge and highlighting the need for more versatile LVLMs.

\begin{table}[tbp]
  \centering
  \caption{The modalities and anatomies involved in our OmniMedVQA dataset.}
    \begin{tabular}{l|p{6cm}}
    \toprule[1pt]
    Modality & Colposcopy, CT (Computed Tomography), Digital Photography, Fundus Photography, Infrared Reflectance Imaging, MR (Magnetic Resonance Imaging), OCT (Optical Coherence Tomography), Dermoscopy, Endoscopy, Microscopy Images, X-Ray, Ultrasound \\
    \hline
    Anatomy   &  Lung, Mammary Gland, Lung, Hand, Upper Limb, Eye, Uterus, Intestine, Skin, Shoulder, Kidney, Gallbladder, Pancreas, Spleen, Liver, Pelvic, Ovary, Blood Vessel, Spine, Urinary System, Adipose Tissue, Muscle Tissue, Oral Cavity, Knee, Foot, Lower Limb    \\
    \bottomrule[1pt]
    \end{tabular}%
  \label{tab:info}%
  \vspace{-6mm}
\end{table}%

\section{Evaluation Method}

\begin{figure}[tbp]
\centering 
\includegraphics[width=1\linewidth]{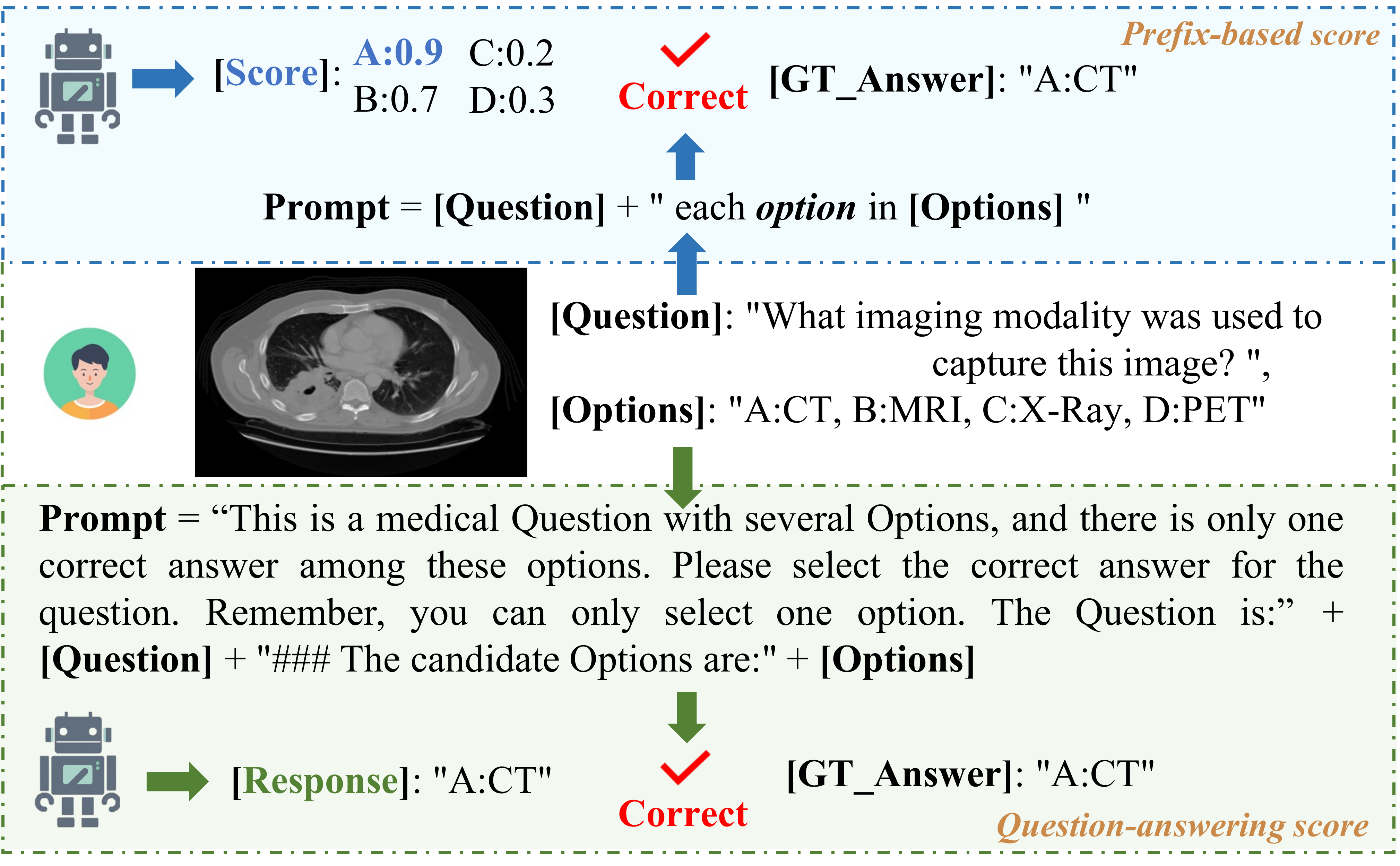} 
\caption{Evaluation process when adopting Question-answering score and Prefix-based score respectively.}
\label{fig:process} 
\end{figure}

As mentioned in Sec.~\ref{sec:data}, for the purpose of evaluation convenience, we provide each QA pair with incorrect answers as candidate options, resulting in a multi-choice Question-Answer task. However, some LVLMs, especially the medical-specialized LVLMs, exhibit poor instruction-following performance during the evaluation, failing to generate responses according to the given options. We think that this situation does not necessarily imply that these models lack medical knowledge, which may simply indicate that they are not proficient in processing input in the form of multiple-choice questions. Therefore, to ensure a more fair comparison, we adopt two different metrics in our evaluation, Question-answering Score and Prefix-based Score \cite{xu2023lvlm, li2023blip}. Their evaluation processes are depicted in Fig.~\ref{fig:process}, and we report their performances respectively in Sec.~\ref{sec:exp}.

\subsection{Question-answering Score}
Given an input image with the question expressions and candidate options, we first combine them to construct the prompt for LVLM. For example, we can utilize the prompt template: ``This is a medical question with several Options, and there is only one correct answer among these options. Please select the correct answer for the question. Remember, you can only select one option. The Question is:\textless{Question}\textgreater{}. \#\#\# The candidate Options are:\textless{Options}\textgreater{}'', based on this template, we insert the current question and candidate options. Then, we deliver the image with the prompt to the LVLM to generate the response. Afterwards, following previous works \cite{xu2023lvlm}, we calculate the similarity of the response with the candidate options and select the option with the largest similarity as the final prediction. Finally, we compare the prediction with the ground-truth answer and judge the correctness.

\subsection{Prefix-based Score}
We also utilize the prefix-based score to evaluate the inherent biomedical knowledge and avoid hallucination in the response. Specifically, in the context of multi-choice Question-Answer tasks, given an input image with the textual sentence, we first extract the visual features and text embeddings, respectively. Then, the visual features are prefixed into the text embeddings, which are subsequently delivered into the LLM to calculate the likelihood score \cite{xu2023lvlm}. This score is considered the prefix-based score for this image-text pair, which reflects the probability of the model generating the corresponding textual content. Therefore, for each candidate option within the specific item, we combine it with the question and then calculate the prefix-based score. The option yielding the highest prefix-based score means it is the most likely answer for this question, which is considered as the final answer of the corresponding LVLM. Then, we compare the final answer with the ground-truth answer to judge the correctness, based on which we compute the VQA accuracy.

It is worthwhile to mention that, although the prefix-based score is not directly equivalent to the response from the LVLM, it measures the likelihood of each option being regarded as the correct answer, which reflects the level of inherent knowledge of the model. In fact, during our evaluation, our primary objective is to find the shortcomings of existing LVLMs in the biomedical field and propose insightful suggestions for future research. However, considering the challenges of OmniMedVQA, it is really hard for these LVLMs to directly generate the correct answer for all the questions. Therefore, the prefix-based score aligns with our evaluation criteria and is a fair metric for the evaluation.

\section{Experiment}
\label{sec:exp}

\subsection{Experimental Details}

In this section, we perform zero-shot evaluation to assess 12 representative LVLMs.  All the experimental environments and hyper-parameters are set according to their released code. Specifically, since MedVInT \cite{zhang2023pmc} has two variants versions that exhibit different capabilities in open-ended and multiple-choice tasks tasks, we employ MedVInT-TD and MedVInT-TE for the evaluation via Question-answering Score and Prefix-based Score, respectively.

\subsection{Overall Performance}
The evaluation results are listed in Table~\ref{tab:resul} which is divided into two sections based on rows. The first eight rows present the accuracy of different general-purpose LVLMs, while the last four rows reflect the performance of medical-domain LVLMs. Specifically, we report the Question-answering Score and Prefix-based Score separately in Table~\ref{tab:resul}. Generally speaking, our OmniMedVQA is extremely challenging, with most LVLMs only slightly surpassing the performance of random guess. Moreover, we have several observations based on the results in Table \ref{tab:resul}. 

\begin{table*}[tbp]
  \centering
  \caption{The accuracy of representative LVLMs on our OmniMedVQA in terms of five different question types. Notably, we report the Question-answering Score and Prefix-based Score before and after ``/'', respectively. Meanwhile, in each column, the best performance is marked in red, while the second best performance is marked in blue.}
  \vspace{-1mm}
    \begin{tabular}{l|c|c|c|c|c|c}
    \toprule[1pt]
    Model & \makecell{Modality \\ Recognition}  & \makecell{Anatomy \\ Identification} & \makecell{Disease \\ Diagnosis} & \makecell{Lesion \\ Grading} & \makecell{Other Biological \\ Attributes} & \makecell{Overall} \\
    \hline
    Random Guess & 25.00 & 25.84 & 28.41 & 25.40 & 37.49 & 28.28\\
    \hline
    MiniGPT-4 \cite{zhu2023minigpt}   &  28.23 / 28.71 & 28.41 / 30.26 & 30.51 / 20.24 & 32.85 / 38.53 & 37.30 / \textcolor{blue}{43.28} & 30.53 / 25.68 \\
    BLIP-2 \cite{li2023blip}  &  57.51 / \textcolor{blue}{37.85} & \textcolor{red}{49.19} / \textcolor{red}{64.75} & \textcolor{red}{46.24} / 23.19 & 30.52 / 25.03 & \textcolor{red}{73.52} / 37.70 & \textcolor{red}{50.69} / \textcolor{red}{33.43} \\
    InstructBLIP \cite{dai2023instructblip}   &  \textcolor{red}{70.62} / 24.83 & \textcolor{blue}{42.75} / \textcolor{blue}{51.78} & 33.62 / 22.41 & \textcolor{red}{54.60} / \textcolor{blue}{54.71} & \textcolor{blue}{48.16} / 28.85 & \textcolor{blue}{42.49} / 28.71   \\
    mPLUG-Owl \cite{ye2023mplug}   &  27.93 / 16.46 & 24.44 / 30.45 & 30.31 / \textcolor{red}{29.32} & 38.50 / \textcolor{red}{76.96} & 37.84 / \textcolor{red}{43.70} & 29.90 / \textcolor{blue}{29.89}  \\
    Otter \cite{li2023otter}   &  25.62 / 10.53 & 25.14 / 23.47 & 27.12 / 22.11 & 34.22 / 39.53 & 32.12 / 24.74 & 27.20 / 21.17  \\
    LLaVA \cite{liu2023visual} &  27.21 / 14.43 & 25.99 / 15.08 & 27.35 / 19.60 & 38.53 / 26.82 & 32.68 / 38.24 & 27.85 / 20.02 \\
    LLaMA\_Adapter\_v2 \cite{gao2023llama}   &   44.51 / 33.48 & 33.73 / 40.67 & 29.17 / 24.93 & \textcolor{blue}{39.07} / 38.65 & 36.80 / 34.07 & 33.15 / 29.88 \\
    VPGTrans \cite{zhang2023transfer}   &  29.32 / 31.77 & 30.76 / 36.27 & 26.91 / 18.88 & 29.61 / 38.53 & 32.65 / 41.47 & 28.49 / 26.15 \\
    \hline
    Med-Flamingo \cite{moor2023med}   &  28.67 / 21.21 & 25.32 / 25.16 & \textcolor{blue}{41.47} / 26.10 & 31.25 / 49.60 & 35.27 / 34.97 & 36.17 / 26.54 \\

    RadFM \cite{wu2023towards}   &   21.31 / \textcolor{red}{38.57} & 19.96 / 29.48 & 28.46 / 25.66 & 24.72 / 35.86 & 37.58 / 31.73 & 26.82 / 29.00 \\

    MedVInT \cite{zhang2023pmc}   &  \textcolor{blue}{59.79} / 30.92 & 41.36 / 23.90 & 36.79 / 25.02 & 15.49 / 5.23 & 46.79 / 30.12 & 41.50 / 25.81 \\

    LLaVA-Med \cite{li2023llava}   &  31.38 / 12.02 & 28.34 / 22.06 & 28.01 / \textcolor{blue}{27.25} & 32.35 / 30.98 & 29.23 / 25.88 & 28.78 / 24.06   \\  

    \bottomrule[1pt]
    \end{tabular}%
  \label{tab:resul}%
\end{table*}%

\begin{table*}[tbp]
  \centering
  \small
  \caption{The overall accuracy of representative LVLMs on our OmniMedVQA in terms of different modalities. Here, we report the accuracy of all items within each modality when utilizing the \textbf{Question-answering score}. Specifically, \textbf{Co} denotes Colposcopy, \textbf{CT} denotes Computed Tomography, \textbf{DP} denotes Digital Photography, \textbf{FP} denotes Fundus Photography, \textbf{IRI} denotes Infrared Reflectance Imaging, \textbf{MR} denotes Magnetic Resonance Imaging, \textbf{OCT} denotes Optical Coherence Tomography, \textbf{Der} denotes Dermoscopy, \textbf{End} denotes Endoscopy, \textbf{Mic} denotes Microscopy Images, \textbf{US} denotes Ultrasound. Meanwhile, in each column, the best and second-best performance are marked in red and blue, respectively.}
  \vspace{-1mm}
    \begin{tabular}{l|c|c|c|c|c|c|c|c|c|c|c|c}
    \toprule[1pt]
    Model & Co & CT & DP & FP & IRI & MR & OCT & Der & End & Mic & X-Ray & US\\
    \hline
    MiniGPT-4 \cite{zhu2023minigpt} & 23.67 & 22.81 & 18.05 & \textcolor{blue}{42.37} & 38.51 & 27.60 & 31.40 & 40.09 & 30.26 & 28.05 & 39.75 & 25.50 \\
    BLIP-2 \cite{li2023blip} & \textcolor{red}{48.52} & \textcolor{red}{56.74} & 23.01 & \textcolor{red}{57.66} & \textcolor{red}{66.18} & \textcolor{blue}{41.77} & \textcolor{red}{68.08} & 41.07 & \textcolor{red}{48.85} & \textcolor{red}{50.17} & \textcolor{red}{70.55} & 37.27 \\
    InstructBLIP \cite{dai2023instructblip}   & 32.25 & 28.72 & \textcolor{blue}{35.75} & 37.72 & \textcolor{blue}{59.27} & 33.79 & 42.59 & \textcolor{red}{61.86} & 36.65 & \textcolor{blue}{48.20} & \textcolor{blue}{61.21} & \textcolor{blue}{41.25} \\
    mPLUG-Owl \cite{ye2023mplug}  & 36.69 & 24.54 & 19.81 & 41.81 & 38.44 & 29.82 & \textcolor{blue}{43.76} & 35.98 & 24.45 & 25.99 & 28.29 & 21.40 \\
    Otter \cite{li2023otter}  & 33.73 & 18.53 & 18.20 & 37.70 & 30.70 & 26.37 & 29.64 & 42.66 & 33.94 & 22.94 & 31.73 & 23.49 \\
    LLaVA \cite{liu2023visual} & 12.72 & 17.73 & 22.25 & 32.16 & 31.23 & 26.99 & 33.73 & 49.67 & 38.20 & 27.95 & 31.35 & 18.66 \\
    LLaMA\_Adapter\_v2 \cite{gao2023llama}  & 38.46 & 21.41 & 28.93 & 36.85 & 35.70 & 27.23 & 33.00 & \textcolor{blue}{51.43} & \textcolor{blue}{46.62} & 34.78 & 46.70 & 34.05 \\
    VPGTrans \cite{zhang2023transfer}  & 32.54 & 21.26 & 20.10 & 34.50 & 32.60 & 25.36 & 25.14 & 44.66 & 30.53 & 23.61 & 46.53 & 25.45 \\
    \hline
    Med-Flamingo \cite{moor2023med} & 18.40 & 38.47 & 21.48 & 27.61 & 39.69 & 40.01 & 26.51 & 32.33 & 30.97 & 46.60 & 28.30 & 24.64 \\
    RadFM \cite{wu2023towards}  & 15.43 & 27.44 & 13.25 & 28.99 & 36.13 & 24.16 & 32.80 & 39.03 & 28.40 & 24.81 & 29.21 & 16.57 \\
    MedVInT \cite{zhang2023pmc}  & \textcolor{blue}{39.17} & \textcolor{blue}{40.74} & \textcolor{red}{43.89} & 39.69 & 46.22 & \textcolor{red}{42.84} & 23.26 & 29.13 & 30.11 & 40.71 & 56.62 & \textcolor{red}{41.26} \\
    LLaVA-Med \cite{li2023llava}  & 28.99 & 18.69 & 18.34 & 35.14 & 30.68 & 27.49 & 34.61 & 44.90 & 41.88 & 26.33 & 31.26 & 29.88 \\
    \bottomrule[1pt]
    \end{tabular}%
  \label{tab:resulmsd}%
\end{table*}%

\begin{table*}[tbp]
  \centering
  \small
  \caption{The overall accuracy of representative LVLMs on our OmniMedVQA in terms of different modalities. Here, we report the accuracy of all items within each modality when utilizing \textbf{Prefix-based score}. Specifically, \textbf{Co} denotes Colposcopy, \textbf{CT} denotes Computed Tomography, \textbf{DP} denotes Digital Photography, \textbf{FP} denotes Fundus Photography, \textbf{IRI} denotes Infrared Reflectance Imaging, \textbf{MR} denotes Magnetic Resonance Imaging, \textbf{OCT} denotes Optical Coherence Tomography, \textbf{Der} denotes Dermoscopy, \textbf{End} denotes Endoscopy, \textbf{Mic} denotes Microscopy Images, \textbf{US} denotes Ultrasound. Meanwhile, in each column, the best and second-best performance are marked in red and blue, respectively.}
  \vspace{-3mm}
    \begin{tabular}{l|c|c|c|c|c|c|c|c|c|c|c|c}
    \toprule[1pt]
    Model & Co & CT & DP & FP & IRI & MR & OCT & Der & End & Mic & X-Ray & US \\
    \hline
    MiniGPT-4 \cite{zhu2023minigpt} & 26.33 & 29.46 & 22.94 & 23.59 & \textcolor{blue}{43.61} & 12.75 & 30.00 & 25.18 & 26.90 & 27.60 & 40.11 & 27.20 \\
    BLIP-2 \cite{li2023blip} & 32.25 & \textcolor{blue}{38.87} & 25.95 & 20.01 & 43.40 & 20.45 & 18.70 & 19.80 & 25.95 & 28.43 & \textcolor{blue}{49.81} & \textcolor{red}{81.79} \\
    InstructBLIP \cite{dai2023instructblip}   & 17.46 & 35.66 & 10.01 & 26.74 & 27.26 & 15.02 & 51.74 & \textcolor{blue}{29.53} & 28.77 & 22.80 & 40.85 & \textcolor{blue}{58.51} \\
    mPLUG-Owl \cite{ye2023mplug}  & 18.64 & 37.00 & 18.31 & \textcolor{red}{56.87} & \textcolor{red}{44.96} & 13.22 & 41.76 & 18.75 & \textcolor{blue}{35.32} & \textcolor{red}{31.04} & 32.18 & 29.14 \\
    Otter \cite{li2023otter}  & 2.37 & 32.55 & 20.78 & 16.96 & 21.90 & 10.73 & 45.98 & 22.52 & 23.39 & 21.99 & 27.80 & 20.88 \\
    LLaVA \cite{liu2023visual}  & 33.73 & 38.27 & 23.55 & 13.84 & 36.37 & 4.65 & 51.23 & 14.17 & 22.16 & 17.65 & 24.10 & 20.74  \\
    LLaMA\_Adapter\_v2 \cite{gao2023llama}  & 28.40 & 36.03 & 21.36 & 22.11 & 33.69 & 17.26 & \textcolor{blue}{54.22} & 21.61 & 32.23 & \textcolor{blue}{30.60} & 41.71 & 47.51 \\
    VPGTrans \cite{zhang2023transfer}  & \textcolor{blue}{35.21} & 30.30 & \textcolor{blue}{27.71} & 23.93 & 42.35 & 11.04 & 31.60 & 22.18 & 24.77 & 25.87 & 39.94 & 40.89 \\
    \hline
    Med-Flamingo \cite{moor2023med} & 30.56 & 22.25 & 7.40 & \textcolor{blue}{51.57} & 30.22 & 14.43 & \textcolor{red}{58.57} & \textcolor{red}{39.48} & \textcolor{red}{46.61} & 23.04 & 38.13 & 17.42 \\
    RadFM \cite{wu2023towards}  & 22.55 & \textcolor{red}{45.47} & 13.43 & 21.48 & 25.16 & 24.52 & 37.40 & 25.71 & 34.44 & 20.52 & \textcolor{red}{55.21} & 24.78 \\
    MedVInT \cite{zhang2023pmc}  & \textcolor{red}{60.24} & 37.77 & \textcolor{red}{39.04} & 9.45 & 32.92 & \textcolor{red}{30.10} & 18.54 & 20.72 & 24.03 & 15.35 & 29.09 & 25.43 \\
    LLaVA-Med \cite{li2023llava}  & 13.61 & 36.75 & 10.48 & 16.15 & 21.61 & \textcolor{blue}{24.58} & 51.51 & 25.35 & 27.92 & 17.40 & 25.54 & 16.75 \\
    \bottomrule[1pt]
    \end{tabular}%
  \label{tab:resulmsl}%
  \vspace{-3mm}
\end{table*}%

\begin{enumerate}	
\item To our surprise, the general-domain LVLM, BLIP2 \cite{li2023blip}, achieves the best performance for all tasks on average, surpassing all tested LVLM models in the medical domain by a large margin. This suggests that emerging property does not occur when using the current medical data to adapt general-purpose LVLMs for medical tasks. 

\item A strong model in aligning image-text pairs in the medical domain is urgently needed for medical-domain LVLMs. BLIP performs well in both general-purpose QA dataset \cite{xu2023lvlm, shao2023tiny} and our OmniMedQA because it is trained by massive high-quality image-text pairs in various visual domains. Hence, the key to developing general medical-purpose LVLMs lies in training models with massive high-quality image captioning data from various medical domains. 

\item MedVInT and Med-Flamingo achieve the highest overall accuracy among all evaluated medical LVLMs and also outperforms many general-purpose LVLMs except for BLIP2 and InstructBLIP. This success may be attributed to the extensive medical knowledge they are injected in the training. Med-Flamingo learns from more than 4k textbooks while MedVInT is trained based on 381K image-caption pairs. This indicates that, to obtain a better performance, more knowledge in the medical domain should be injected into the LVLMs.

\item Through a comparative analysis of LLaVA and LLaVA-Med, we conclude that medical instruction tuning can improve the performance of general-purpose LVLM in the biomedical field. However, when comparing different medical LVLMs, we can find that LLaVA-Med delivers the worst performance. As listed in Table~\ref{tab:comparison-LVLMs}, LLaVA-Med initiates its model from the pre-trained LLaVA \cite{li2023llava} and incorporates only a small amount of data in the training, through which they expect to save the training cost. However, the evaluation results suggest the LLaVA pre-trained model is not suitable and the advanced performance should be instructed by sufficient data, which calls for a robust pre-trained model tailored to the biomedical field and high-quality data. In fact, the success of MiniGPT-4 in the general domain demonstrates that high-quality data, even if it is in small quantities, could strongly support the training of LVLM. However, the instruction data for LLaVA-Med is generated through the GPT-4 API and only relies on textual information. We believe this substantially compromises data quality, leading to the diminished accuracy. This underscores the need for an effective caption-generation model like BLIP2 \cite{li2023blip}, which can generate accurate and detailed textual descriptions for biomedical images.

\end{enumerate}

Overall, through our evaluation, we find that medical-specialized LVLMs do not present outstanding performance. For most question types, two general LVLMs, BLIP2 and InstructBLIP obtain the best accuracy. To better elaborate the underlying reasons, we conduct an in-depth analysis of the evaluation results in Sec.~\ref{sec:ana}.

\subsection{Analysis in terms of modalities}
\label{sec:ana}
To further analyze the performance of LVLMs in the biomedical field, we report the accuracy of all QA items in terms of different modalities in Table~\ref{tab:resulmsd} and Table~\ref{tab:resulmsl}, which present the Question-answering score and Prefix-based score respectively. Based on the results, we have following two observations. 

\begin{enumerate}

\item Although medical LVLMs exhibit lower accuracy when considering the overall dataset, they tend to perform well in modalities characterized by substantial differences from general images, such as CT and MRI. However, in modalities with similar distributions to those in general domain images, medical-specialized LVLMs fail to demonstrate notably superior performance.

\item RadFM aims to initiate the development of the radiology foundation model, which is trained with more than 19M radiologic image-test pairs. Among their dataset, CT, MRI and X-Ray constitute a significant proportion. Therefore, as listed in Table~\ref{tab:resulmsl}, RadFM achieves the best performance on CT and X-Ray tasks, and obtain competitive performance on MR task. This reveals the potential for performance improvement with high-quality instruction data. To bring an all-around LVLM in the biomedical field, the inclusion of additional high-quality data from various modalities, such as Fundus Photography and Infrared Reflectance Imaging, is imperative.

\end{enumerate}
\section{Conclusion}
Since LVLMs have recently received remarkable attention in the whole community, this paper aims to evaluate their performance in the biomedical field. To achieve this goal, we collect OmniMedVQA, a large-scale medical VQA dataset. OmniMedVQA has 118,010 images with 127,995 question-answer items, which include 12 different modalities and cover more than 20 human anatomical regions. Therefore, OmniMedVQA could support the throughout evaluation of different LVLMs. During the evaluation, we assess 12 different LVLMs, including 8 general-domain models and 4 specialized LVLMs for the biomedical field. To our great surprise, despite their claims of robustness, the medical LVLMs exhibit inferior performance to those general-domain models, which reveals the shortcomings of these medical models. We point out that to become a more versatile medical expert, medical LVLMs consistently require additional knowledge of specific modalities, such as Infrared Reflectance Imaging and Fundus Photography, on which the performance of medical LVLMs is significantly inferior to general-domain models. We hope our dataset provides a comprehensive evaluation benchmark for medical LVLMs and our findings offer useful suggestions for future research.

\section*{Acknowledgement}
This paper is partially supported by the National Key R\&D Program of China No.2022ZD0161000 and the General Research Fund of Hong Kong No.17200622.
\clearpage
\setcounter{page}{1}
\maketitlesupplementary

\section{The details of involved datasets}
To construct comprehensive evaluation benchmark, we collect numerous medical datasets and convert them into VQA format. In Table~\ref{tab:alldatasets1} and Table~\ref{tab:alldatasets2}, we provide a list of all the datasets included in OmniMedVQA, along with their modality information, the number of utilized images and QA items, and the access condition. It's worth noting that RadImageNet \cite{mei2022radimagenet} stands out as one of the largest datasets in the biomedical field, containing 1.35 million radiologic images encompassing CT and MRI modalities, spanning 11 anatomical regions, and covering 165 different diseases. As a result, RadImageNet constitutes a significant portion of our OmniMedVQA dataset.

Additionally, 3D\_Modality is our self-construction dataset, incorporating data from 17 different medical datasets. The involved 17 datasets are ISLES\_SPES\cite{maier2017isles}, ISLES2016\cite{winzeck2018isles}, ISLES2017\cite{winzeck2018isles}, ISLES2018\cite{cereda2016benchmarking, hakim2021predicting}, ISLES2022\cite{hernandez2022isles}, AMOS\cite{ji2022amos}, Longitudinal Multiple Sclerosis Lesion Segmentation\cite{carass2017longitudinal},VALDO\cite{VALDO_Task2}, PICAI\cite{saha2021end}, ASC18\cite{xiong2021global}, BraTS2013\cite{menze2014multimodal, kistler2013virtual}, BraTS2018\cite{menze2014multimodal,bakas2017advancing,bakas2018identifying}, MSSEG2016\cite{commowick2021multiple}, CMRxMotions\cite{wang2022extreme}, MRBrainS13\cite{mendrik2015mrbrains}, BrainTumour\cite{antonelli2022medical}, MRBrain18\cite{kuijf2019grand}. We leverage these datasets to create questions about more fine-grained modality recognition, such as Magnetic Resonance T1-weighted, Magnetic Resonance T2-weighted, Magnetic Resonance T1-weighted Inversion Recovery, Computed Tomography Cerebral Blood Volume, Computed Tomography Time to Maximum \emph{et al.}

We release the OmniMedVQA according to the license and permission. Specifically, there are 42 dataset are completely open access. Thus, we directly provide the images with the corresponding QA items. Meanwhile, there are 31 datasets are restricted access. For these datasets, we only release the evaluation QA items and provide the instruction guidelines, based on which you can associate each QA item with the corresponding images. Researchers only need to download the original datasets and then combine them with the provided QA items according to the guidelines.

For the convenience of future research, in Table~\ref{tab:resul_open-access}, Table~\ref{tab:resulmsd_open-access} and Table~\ref{tab:resulmsl_open-access}, we provide the evaluation results on the completely open-access datasets. If you do not want to download each restricted access dataset one by one, these results could help you establish the benchmark and analyse the experimental results quickly.

\section{The distribution of our dataset}
\begin{figure*}[tbp]
\centering 
\includegraphics[width=0.8\linewidth]{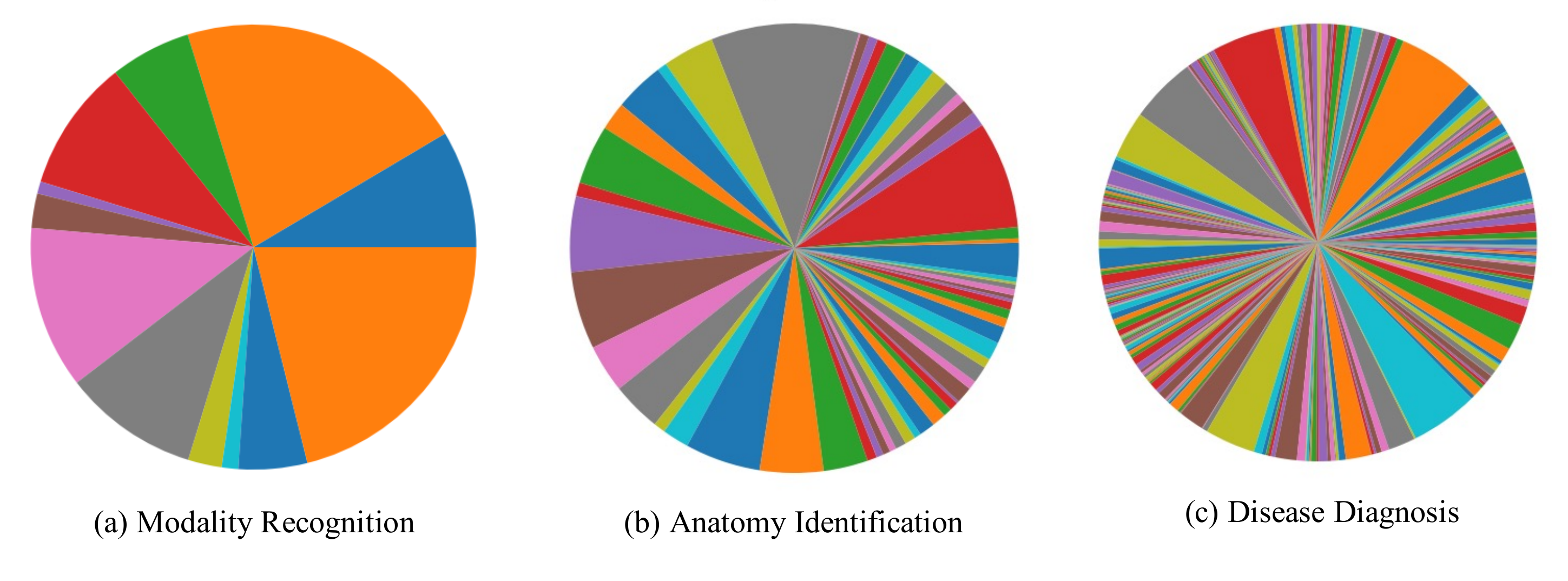} 
\caption{We illustrate the distribution of different classes within Modality Recognition, Anatomy Identification and Disease Diagnosis.}
\label{fig:pie_chart1} 
\end{figure*}

We illustrate the distribution of different classes within Modality Recognition, Anatomy Identification and Disease Diagnosis in Fig.~\ref{fig:pie_chart1}. We can find there is no significant bias in our OmniMedVQA and the distribution remains balanced. This demonstrates the effectiveness of the sampling process when we develop the dataset.

\section{The details of modalities}
In this section, we provide an overview of the number of images and QA items associated with various modalities in OmniMedVQA. As detailed in Table~\ref{tab:mosta}, we incorporate data from 12 different modalities. Moreover, to better present the characteristic of different modalities, we illustrate the images with the corresponding QA items in Fig.~\ref{fig:modality1} and Fig.~\ref{fig:modality2}.

\section{The details of multi-choice questions}
\begin{figure*}[tbp]
\centering 
\includegraphics[width=1\linewidth]{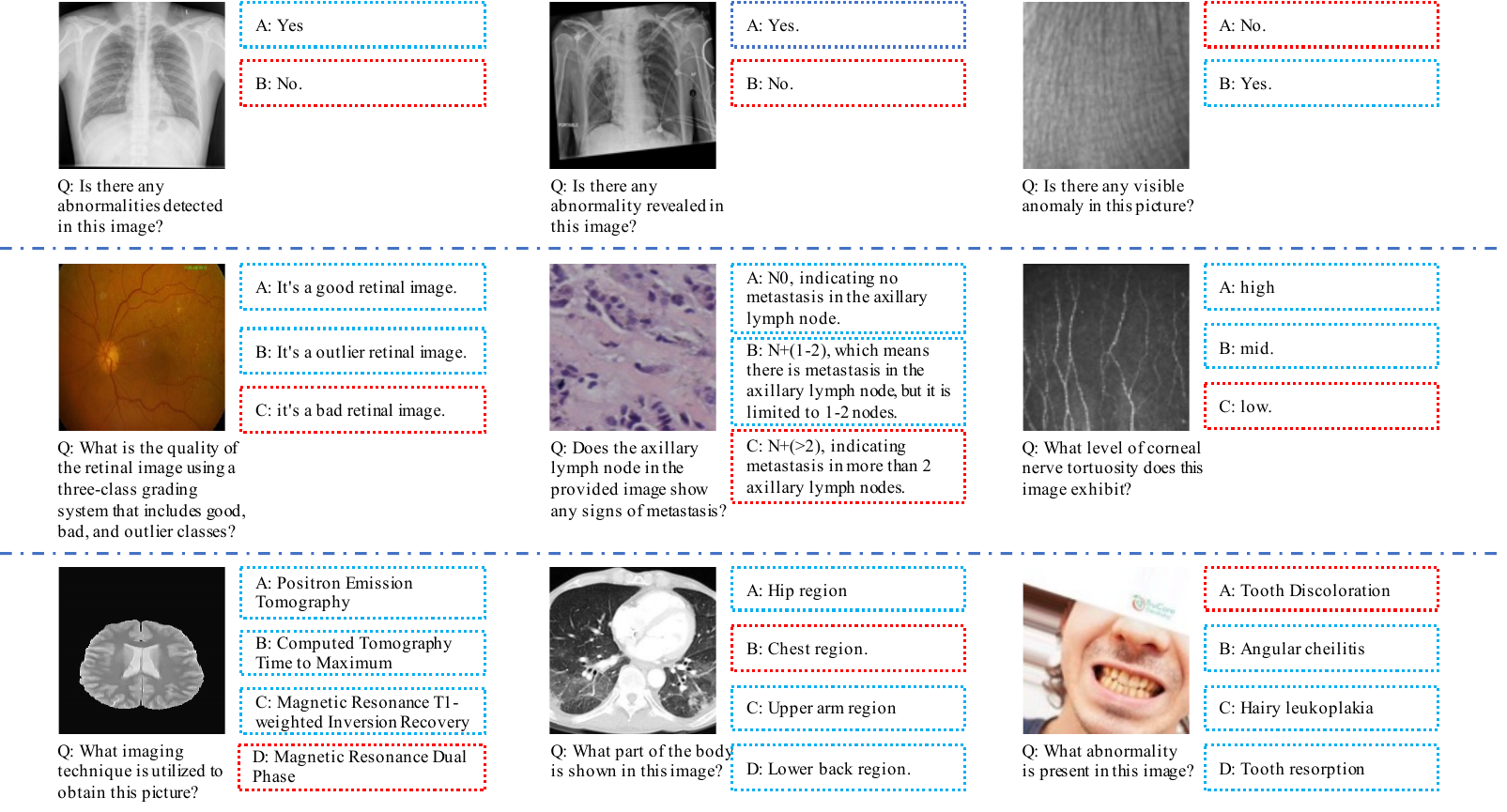} 
\caption{Illustration of representative samples with different numbers of candidate options.}
\label{fig:option} 
\end{figure*}

As introduced in Sec~\ref{sec:data}, we generate a set of incorrect options for each item, which are utilized to construct multiple-choice question-answer pairs. The number of candidate options of each question ranges from 2 to 4. In Fig.~\ref{fig:option}, we illustrate the QA items with different number of options. As depicted, questions with two options are ``Yes/No'' selection. On the other hand, questions with three options predominantly focus on Lesion Grading, which judges the severity of the disease.

\begin{table*}[tbp]
  \centering
  \footnotesize
  \caption{The information of involved dataset in OmniMedVQA. Notably, \textbf{Co} denotes Colposcopy, \textbf{CT} denotes Computed Tomography, \textbf{DP} denotes Digital Photography, \textbf{FP} denotes Fundus Photography, \textbf{IRI} denotes Infrared Reflectance Imaging, \textbf{MR} denotes Magnetic Resonance Imaging, \textbf{OCT} denotes Optical Coherence Tomography, \textbf{Der} denotes Dermoscopy, \textbf{End} denotes Endoscopy, \textbf{Mic} denotes Microscopy Images, \textbf{US} denotes Ultrasound.}
    \begin{tabular}{c|l|ccc|c}
    \toprule[1pt]
    index & Dataset & Modality  & \# Imgs & \# QA Items & Access  \\
    \hline
    1& TCB\_Challenge \cite{harrar2014texture} & X-Ray & 18 & 32 & Restricted Access \\\hline
    2& Oral\_Cancer\_kaggle \cite{digital_photography_OralCancer} & DP & 27 & 34 & Restricted Access \\\hline
    3& Dental\_Condition\_Dataset \cite{digital_photography_Oral_Diseases} & DP & 2281 & 2752 & Restricted Access \\\hline
    4& Cervical\_Cancer\_Screening \cite{beno2017intel} & Co & 319 & 338 & Restricted Access \\\hline
    5& Chest\_CT\_Scan \cite{ct_2d_chest_ct_scan} & CT & 382 & 871 & Open Access \\\hline
    6& Covid\_CT \cite{ct_2d_covid_ct_covid_ct} & CT & 135& 199 & Open Access \\\hline
    7& SARS-CoV-2 CT-scan \cite{soares2020sars} & CT & 461& 910& Open Access \\\hline
    8& RadImageNet \cite{mei2022radimagenet} & CT,MR,US & 55443 & 56697 & Open Access \\\hline
    9& Fitzpatrick 17k \cite{groh2021evaluating, groh2022towards} & Der & 1450& 1552& Open Access \\\hline
    10& ISBI2016 \cite{gutman2016skin} & Der & 348 & 681 &Open Access \\\hline
    11& ISIC2018 \cite{codella2019skin, tschandl2018ham10000} & Der & 185 & 272 & Open Access \\\hline
    12& ISIC2019 \cite{dermoscopy_isic2019} & Der & 1860 & 1952 & Open Access \\\hline
    13& ISIC2020 \cite{rotemberg2021patient} & Der & 1499 & 1580 & Open Access \\\hline
    14& MED-NODE \cite{giotis2015med} & Der & 34& 38 & Restricted Access \\\hline
    15& Monkeypox Skin Image 2022 \cite{islam2022web} &Der  & 154 & 163 & Open Access \\\hline
    16& PAD-UFES-20 \cite{pacheco2020impact} & Der & 401& 479 & Open Access \\\hline
    17& $PH^{2}$ \cite{mendoncca2013ph} & Der & 36 & 45 & Restricted Access \\\hline
    18& AIDA \cite{endoscopy_aida_e} & End & 207 & 340 & Restricted Access \\\hline
    19& Kvasir \cite{pogorelov2017kvasir} & End & 1225& 1537 & Restricted Access \\\hline
    20& ACRIMA \cite{diaz2019cnns} & FP  & 129 & 159 & Open Access \\\hline
    21& Adam Challenge \cite{fang2022adam} & End &78 &87 & Open Access \\\hline
    22& AIROGS \cite{de2023airogs} & FP  & 3853 & 4004 &Restricted Access \\\hline
    23& APTOS2019\_Blindness \cite{karthik_2019} & FP  & 544 & 625 & Restricted Access \\\hline
    24& AVN Assessment \cite{AVNickingQuantification} & FP &18 &22 & Restricted Access \\\hline
    25& DeepDRiD \cite{liu2022deepdrid} & FP  & 131 & 131 & Open Access \\\hline
    26& Diabetic Retinopathy \cite{diabetic-retinopathy-arranged} & FP  & 1996 & 2051 & Open Access \\\hline
    27& DRIMDB \cite{fundus_photography_drimdb} & FP &122 &132 & Open Access \\\hline 
    28& GAMMA \cite{fundus_photography_gamma} & FP  & 20 & 20 & Restricted Access \\\hline
    29& Glaucoma\_Detection \cite{fundus_photography_glaucoma_detection} & FP & 121 &142 & Restricted Access \\\hline
    30& JSIEC \cite{cen2021automatic} & FP  & 177 & 220 & Open Access \\\hline
    31& Messidor-2 \cite{decenciere2014feedback, abramoff2013automated} & FP  & 270 & 321 & Restricted Access \\\hline
    32& OLIVES \cite{prabhushankar2022olives} & FP &534 & 593 & Open Access \\\hline
    33& PALM2019 \cite{fang2023palm} & FP  & 451 & 510 & Open Access \\\hline
    34& Refuge2 \cite{orlando2020refuge, li2020development} & FP  & 128 & 145 & Restricted Access \\\hline 
    35& Cataract\_dataset\_kaggle \cite{fundus_photography_retina_cataract_dataset} & FP &120 &138 & Restricted Access \\\hline
    36& Yangxi  \cite{liu2019self} & FP  & 1446 & 1515 & Open Access \\\hline
    37& BCNB \cite{xu2021predicting} & MR  & 4334 & 4806 & Restricted Access \\\hline
    38& BRIGHT Challenge \cite{histopathology_bracs} & MR &675 & 890 & Restricted Access \\\hline
    39& BreakHis \cite{spanhol2015dataset} & MR  & 684 & 735 & Open Access \\\hline
    40& NLM- Malaria Data \cite{histopathology_cell_images_malaria} & MR  & 67 & 75 & Open Access \\\hline
    41& CRC100k \cite{kather2018100} & MR &1186 &1322& Open Access \\\hline
    42& DigestPath19 \cite{da2022digestpath} & MR  & 81 & 95 & Restricted Access \\\hline
    43& His\_Can\_Det \cite{cukierski2018histopathologic} & MR  & 7381 & 7572 & Restricted Access \\\hline
    44& lc25000 \cite{borkowski2019lung} & MR &1796 & 1903 & Restricted Access \\\hline
    45& MAlig\_Lymph \cite{orlov2010automatic} & MR  & 75 & 149 & Open Access \\\hline
    46& MRL\_Eye \cite{fusek2018pupil} & IRI  & 9477 & 9785 & Restricted Access \\\hline
    47& BioMediTech \cite{nanni2016texture} & Mic &345 &511& Open Access \\\hline
    48& Blood\_Cell \cite{microscopy_images_blood_cell_images_dataset} & Mic  & 1092 & 1175 & Open Access \\\hline
    49& CornealNerve \cite{scarpa2008automatic} & Mic  & 18 & 25 & Restricted Access \\\hline
    50& Cervix93 \cite{phoulady2018new} & Mic & 434 & 664 & Restricted Access \\\hline
    51& HuSHeM \cite{shaker2017dictionary} & Mic  & 41 & 89 & Open Access \\\hline
    52& BACH2018 \cite{aresta2019bach} & Mic  & 80 & 102 & Restricted Access \\\hline
    53& ALL Challenge \cite{gupta2019isbi} & Mic &295 &342& Open Access \\\hline
    54& MHSMA \cite{javadi2019novel} & Mic  & 1196 & 1282 & Open Access \\\hline
    55& Nerve\_Tortuosity \cite{scarpa2011automatic} & Mic  & 5 & 6 & Restricted Access \\\hline
    56& Br35h \cite{hamada2020br35h} & MR &382 &429 & Restricted Access \\\hline
    57& OCT \& X-Ray 2017 \cite{kermany2018identifying} & OCT,X-Ray  & 1066 & 1301 & Open Access \\\hline
    58& Retinal OCT-C8 \cite{oct_Retina-OCT-C8} & OCT  & 3224 & 4016 & Open Access \\\hline
    59& Knee\_Osteoarthritis \cite{chen2018knee} & X-Ray &518 & 518 & Open Access \\\hline
    60& RUS\_CHN \cite{x_ray_RUS_CHN} & X-Ray  & 1642 & 1982 & Open Access \\\hline
    \end{tabular}%
  \label{tab:alldatasets1}%
\end{table*}%
    
\begin{table*}[tbp]
  \centering
  \footnotesize
  \caption{Continued from Table~\ref{tab:alldatasets1}. The information of involved dataset in OmniMedVQA. Notably, \textbf{Co} denotes Colposcopy, \textbf{CT} denotes Computed Tomography, \textbf{DP} denotes Digital Photography, \textbf{FP} denotes Fundus Photography, \textbf{IRI} denotes Infrared Reflectance Imaging, \textbf{MR} denotes Magnetic Resonance Imaging, \textbf{OCT} denotes Optical Coherence Tomography, \textbf{Der} denotes Dermoscopy, \textbf{End} denotes Endoscopy, \textbf{Mic} denotes Microscopy Images, \textbf{US} denotes Ultrasound.}
    \begin{tabular}{c|l|ccc|c}
    \toprule[1pt]
    index & Dataset & Modality  & \# Imgs & \# QA Items & Access  \\
    \hline
    61& Pulmonary\_Chest\_Shenzhen \cite{jaeger2014two} & X-Ray &131 & 296 & Open Access \\\hline
    62& Chest\_X-Ray\_PA \cite{asraf2021covid19} & X-Ray &664 &850 & Open Access \\\hline
    63& CoronaHack \cite{cohen2020covid} & X-Ray  & 476 & 684& Open Access \\\hline
    64& Covid-19\_tianchi \cite{x_ray_covid_19_image_dataset} & X-Ray  & 66 & 96 & Open Access \\\hline
    65& Covid19\_heywhale \cite{chowdhury2020can} & X-Ray & 550 &690& Open Access \\\hline
    66& COVIDGR \cite{tabik2020covidgr} & X-Ray  & 156 & 220 & Restricted Access \\\hline
    67& COVIDx CXR-4 \cite{wang2020covid} & X-Ray  & 335 & 485 & Open Access \\\hline
    68& MINIJSRT \cite{shiraishi2000development} & X-Ray &133 &257 & Restricted Access \\\hline
    69& MIAS \cite{suckling1994mammographic} & X-Ray  & 65 & 142 & Open Access \\\hline
    70& Mura \cite{rajpurkar2017mura} & X-Ray  & 1277 & 1464 & Open Access \\\hline
    71& Pulmonary\_Chest\_MC \cite{jaeger2014two} & X-Ray &28 &38 & Open Access \\\hline
    72& SIIM-ACR \cite{zawacki2019siim} & X-Ray  & 1036 & 1286 & Restricted Access \\\hline
    73& 3D\_Modality & MR,CT  & 426 & 426 & Partially-Open Access \\\hline
    \end{tabular}%
  \label{tab:alldatasets2}%
\end{table*}%

\begin{figure*}[t]
\centering 
\includegraphics[width=1\linewidth]{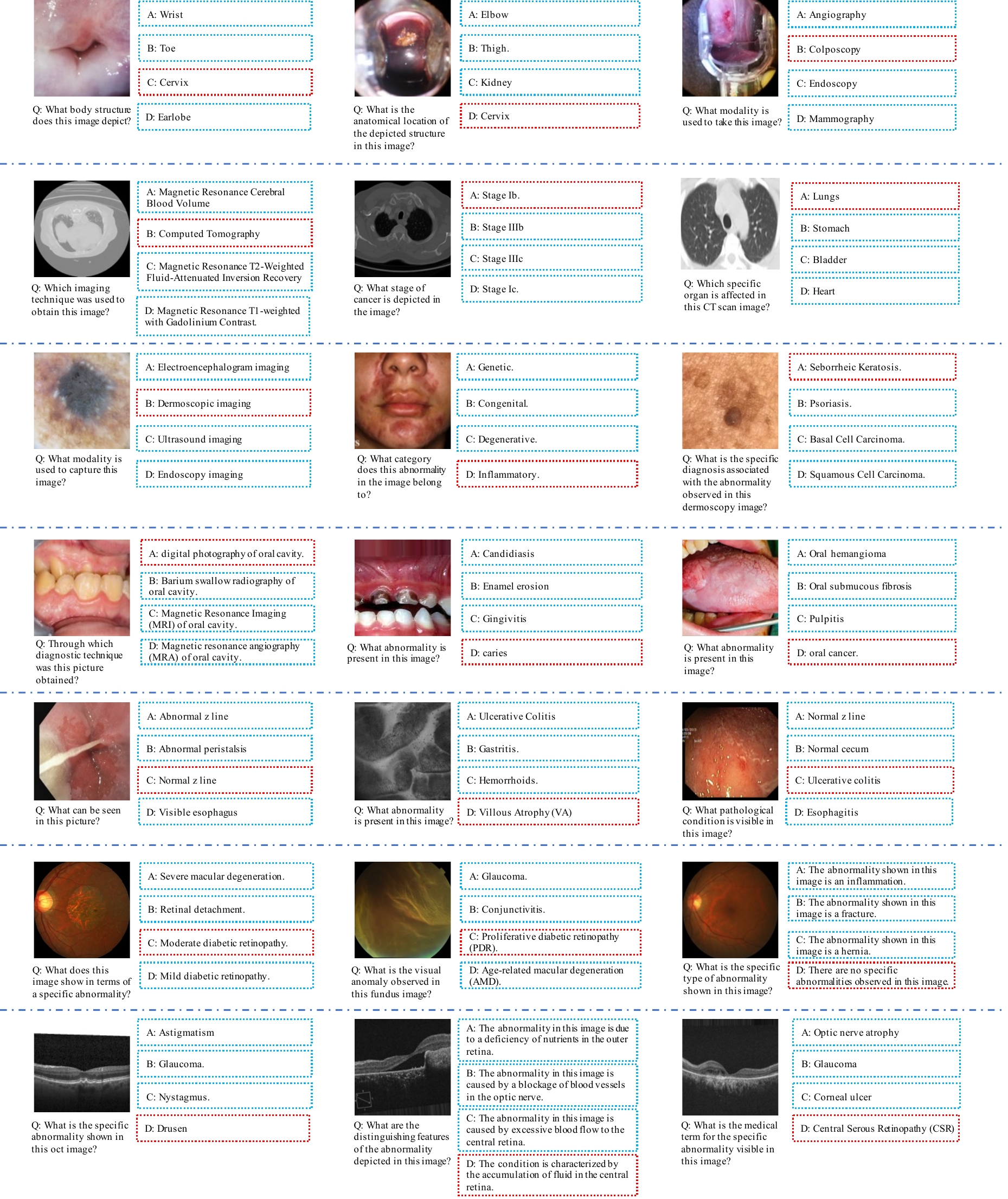} 
\caption{The representative samples from different modalities. From the above to bottom, we illustrate the samples from 7 different modalities in each row, \emph{i.e.,} Colposcopy, CT, Dermoscopy, Digital Photography, Endoscopy, Fundus Photography and OCT. Notably, each dashed box corresponds to a specific option, and the red dashed box indicates the correct option.}
\label{fig:modality1} 
\end{figure*}

\begin{figure*}[tbp]
\centering 
\includegraphics[width=1\linewidth]{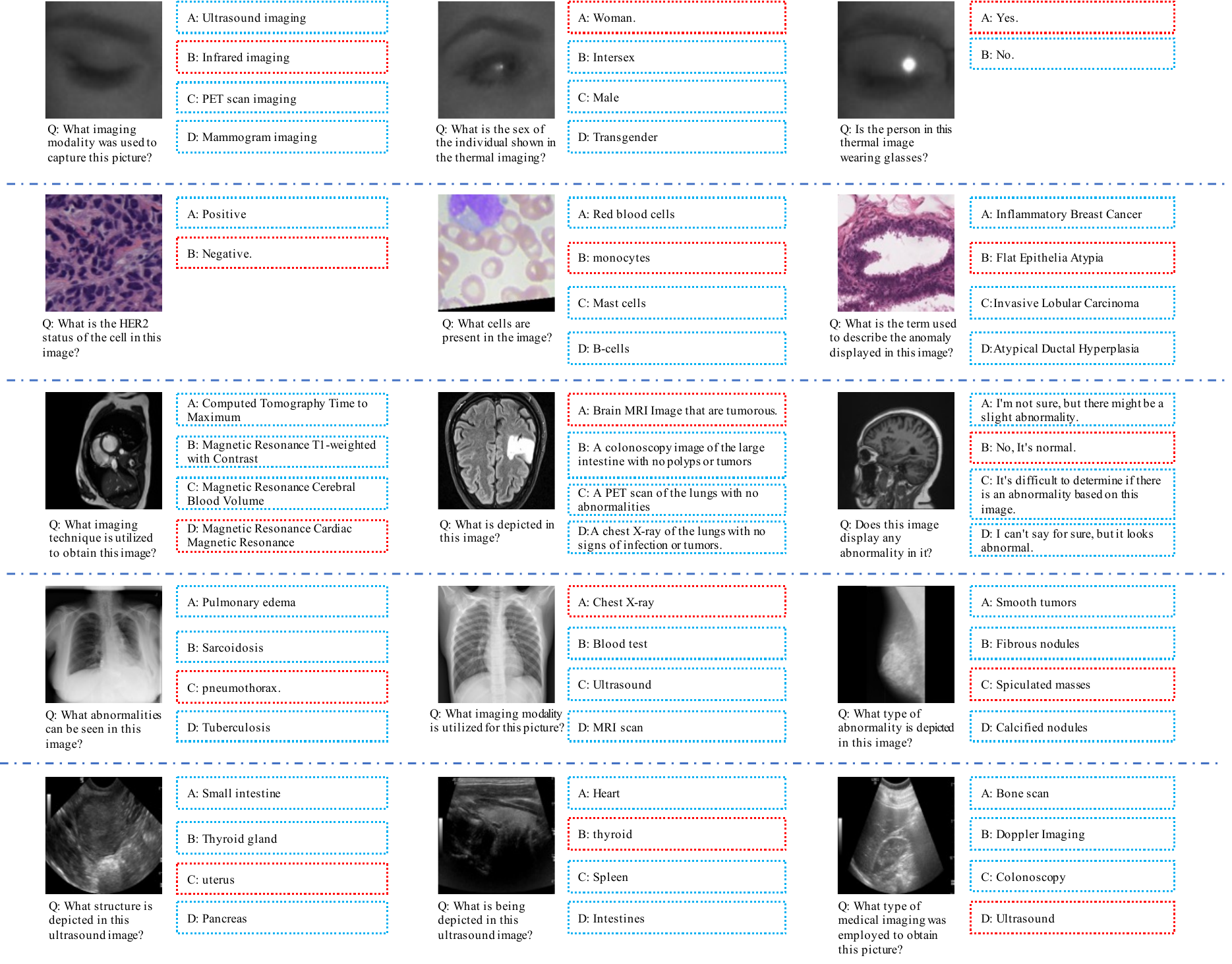} 
\caption{Continued from Fig.~\ref{fig:modality1}. The representative samples from different modalities. From the above to bottom, we illustrate the samples from 5 different modalities in each row, \emph{i.e.,} Infrared Reflectance Imaging, Microscopy Images, MR, X-Ray and Ultrasound. Notably, each dashed box corresponds to a specific option, and the red dashed box indicates the correct option.}
\label{fig:modality2} 
\end{figure*}

\begin{table}[tbp]
  \centering
  \caption{The numbers of images and QA items sourced from different modalities in our OmniMedVQA.}
    \begin{tabular}{l|c|c}
    \toprule[1pt]
    Modality &  \# Images & \# QA Items  \\
    \hline
    Colposcopy    &  319 &  338   \\
    CT  &  14457 &  15836  \\
    Digital Photography    & 2308 &  2786 \\
    Fundus Photography   & 10108 &  10815 \\
    Infrared Reflectance Imaging   & 9477 &  9785 \\
    MR   & 31917 &  32705 \\
    Optical Coherence Tomography   & 3791 &  4646 \\
    Dermoscopy   & 5967 &  6762 \\
    Endoscopy   & 1432 &  1877 \\
    Microscopy Images   & 19785 &  21743 \\
    X-Ray   & 7594 &  9711 \\
    Ultrasound   & 10855 &  10991 \\
    \bottomrule[1pt]
    \end{tabular}%
  \label{tab:mosta}%
\end{table}%

\begin{table*}[tbp]
  \centering
  \caption{The accuracy of representative LVLMs on completely open-access data of our OmniMedVQA in terms of five different question types. Notably, we report the Question-answering Score and Prefix-based Score before and after ``/'', respectively. Meanwhile, in each column, the best performance is marked in red, while the second best performance is marked in blue.}
    \begin{tabular}{l|c|c|c|c|c|c}
    \toprule[1pt]
    Model & \makecell{Modality \\ Recognition}  & \makecell{Anatomy \\ Identification} & \makecell{Disease \\ Diagnosis} & \makecell{Lesion \\ Grading} & \makecell{Other Biological \\ Attributes} & \makecell{Overall} \\
    \hline
    Random Guess & 25.00 & 25.67 & 25.12 & 27.86 & 25.48 & 26.91 \\
    \hline
    MiniGPT-4 \cite{zhu2023minigpt}   &  26.43 / 25.85 & 28.88 / 30.19 & 30.47 / 19.31 & 34.56 / 40.04 & 30.36 / 39.25 & 29.74 / 23.44  \\

    BLIP-2 \cite{li2023blip}  &  \textcolor{blue}{68.19} / \textcolor{blue}{46.75} & \textcolor{red}{44.39} / \textcolor{red}{72.43} & \textcolor{red}{44.51} / 23.64 & 29.03 / 24.31 & \textcolor{red}{67.95} / 33.85 & \textcolor{red}{48.12} / \textcolor{red}{36.08} \\

    InstructBLIP \cite{dai2023instructblip}   &  \textcolor{red}{75.27} / 24.15 & \textcolor{blue}{44.35} / \textcolor{blue}{57.56} & 32.29 / 24.97 & \textcolor{red}{59.25} / \textcolor{blue}{56.91} & 23.72 / 36.65 & \textcolor{blue}{40.40} / 32.10  \\

    mPLUG-Owl \cite{ye2023mplug}   &  28.95 / 9.30 & 24.83 / 32.18 & 30.13 / 25.39 & 43.61 / \textcolor{red}{86.84} & 28.62 / 34.25 & 29.25 / 26.35  \\

    Otter \cite{li2023otter}   &  24.50 / 7.96 & 25.81 / 25.24 & 27.74 / 22.73 & 37.37 / 41.71 & 26.33 / 28.07 & 27.13 / 21.93  \\

    LLaVA \cite{liu2023visual} &  21.36 / 14.94 & 25.86 / 15.16 & 29.10 / 20.50 & \textcolor{blue}{43.95} / 24.07 & 31.90 / 36.11 & 27.96 / 19.49  \\

    LLaMA\_Adapter\_v2 \cite{gao2023llama}   &  37.29 / 40.36 & 33.72 / 43.30 & 31.19 / 23.52 & 41.99 / 37.13 & 34.22 / \textcolor{red}{41.25} & 32.82 / 30.38 \\

    VPGTrans \cite{zhang2023transfer}   &  26.80 / 31.81 & 31.06 / 37.95 & 30.05 / 17.62 & 30.60 / 40.75 & 29.67 / 39.31 & 29.81 / 24.62 \\
    \hline

    Med-Flamingo \cite{moor2023med}   &  30.19 / 24.25 & 24.93 / 25.57 & \textcolor{blue}{38.90} / 24.29 & 30.74 / 52.48 & 14.18 / 38.25 & 34.03 / 25.73 \\

    RadFM \cite{wu2023towards}   &  13.31 / \textcolor{red}{51.76} & 21.69 / 30.11 & 30.35 / \textcolor{blue}{28.30} & 26.64 / 33.56 & \textcolor{blue}{43.85} / \textcolor{blue}{40.28} & 26.99 / \textcolor{blue}{32.27}  \\

    MedVInT \cite{zhang2023pmc}   &  68.10 / 33.61 & 40.26 / 23.27 & 35.78 / 28.29 & 12.77 / 5.29 & 30.30 / 23.58 & 40.04 / 27.32 \\

    LLaVA-Med \cite{li2023llava}   & 26.93 / 13.06 & 29.53 / 22.47 & 29.22 / \textcolor{red}{32.50} & 34.18 / 30.41 & 33.08 / 22.70 & 29.25 / 27.69   \\  

    \bottomrule[1pt]
    \end{tabular}%
  \label{tab:resul_open-access}%
\end{table*}%

\begin{table*}[tbp]
  \centering
  \small
  \caption{The overall accuracy of representative LVLMs on completely open-access data of our OmniMedVQA in terms of different modalities . Here, we report the accuracy of all items within each modality when utilizing the \textbf{Question-answering score}. Specifically, \textbf{Co} denotes Colposcopy, \textbf{CT} denotes Computed Tomography, \textbf{DP} denotes Digital Photography, \textbf{FP} denotes Fundus Photography, \textbf{IRI} denotes Infrared Reflectance Imaging, \textbf{MR} denotes Magnetic Resonance Imaging, \textbf{OCT} denotes Optical Coherence Tomography, \textbf{Der} denotes Dermoscopy, \textbf{End} denotes Endoscopy, \textbf{Mic} denotes Microscopy Images, \textbf{US} denotes Ultrasound. Meanwhile, in each column, the best and second-best performance are marked in red and blue, respectively.}
    \begin{tabular}{l|c|c|c|c|c|c|c|c|c|c|c|c}
    \toprule[1pt]
    Model & Co & CT & DP & FP & IRI & MR & OCT & Der & End & Mic & X-Ray & US\\
    \hline

    MiniGPT-4 \cite{zhu2023minigpt} & - & 22.81 & - & 38.33 & - & 27.48 & 31.40 & 40.25 & - & 36.23 & 38.30 & 25.50 \\

    BLIP-2 \cite{li2023blip} & - & \textcolor{red}{56.74} & - & 46.24 & - & \textcolor{blue}{41.32} & \textcolor{red}{68.08} & 40.65 & - & \textcolor{red}{50.40} & \textcolor{red}{67.58} & 37.27 \\

    InstructBLIP \cite{dai2023instructblip}   & - & 28.72 & - & \textcolor{blue}{50.31} & - & 33.15 & 42.59 & \textcolor{red}{62.22} & - & \textcolor{blue}{46.29} & \textcolor{blue}{61.04} & \textcolor{blue}{41.25} \\

    mPLUG-Owl \cite{ye2023mplug}  & - & 24.54 & - & 36.92 & - & 29.90 & \textcolor{blue}{43.76} & 36.10 & - & 27.25 & 28.92 & 21.40  \\

    Otter \cite{li2023otter}  & - & 18.53 & - & 37.51 & - & 26.06 & 29.64 & 42.64 & - & 27.48 & 31.85 & 23.49  \\

    LLaVA \cite{liu2023visual} & - & 17.73 & - & 47.11 & - & 26.72 & 33.73 & 49.74 & - & 28.87 & 30.70 & 18.66  \\

    LLaMA\_Adapter\_v2 \cite{gao2023llama}  & - & 21.41 & - & \textcolor{red}{50.74} & - & 26.63 & 33.00 & \textcolor{blue}{51.76} & - & 38.66 & 46.44 & 34.05  \\

    VPGTrans \cite{zhang2023transfer}  & - & 21.26 & - & 45.02 & - & 25.44 & 25.14 & 45.01 & - & 34.70 & 46.64 & 25.45  \\
    \hline

    Med-Flamingo \cite{moor2023med} & - & 38.47 & - & 30.12 & - & 40.56 & 26.51 & 32.43 & - & 19.93 & 30.34 & 24.64 \\

    RadFM \cite{wu2023towards}  & - & 27.56 & - & 36.89 & - & 24.06 & 32.80 & 39.21 & - & 27.96 & 30.95 & 16.57 \\

    MedVInT \cite{zhang2023pmc}  & - & \textcolor{blue}{40.74} & - & 31.84 & - & \textcolor{red}{43.10} & 23.26 & 29.11 & - & 32.00 & 55.10 & \textcolor{red}{41.26} \\

    LLaVA-Med \cite{li2023llava}  & - & 18.69 & - & 39.03 & - & 27.47 & 34.61 & 44.95 & - & 33.29 & 30.68 & 29.88  \\
    \bottomrule[1pt]
    \end{tabular}%
  \label{tab:resulmsd_open-access}%
\end{table*}%

\begin{table*}[tbp]
  \centering
  \small
  \caption{The overall accuracy of representative LVLMs on completely open-access data of our OmniMedVQA in terms of different modalities. Here, we report the accuracy of all items within each modality when utilizing \textbf{Prefix-based score}. Specifically, \textbf{Co} denotes Colposcopy, \textbf{CT} denotes Computed Tomography, \textbf{DP} denotes Digital Photography, \textbf{FP} denotes Fundus Photography, \textbf{IRI} denotes Infrared Reflectance Imaging, \textbf{MR} denotes Magnetic Resonance Imaging, \textbf{OCT} denotes Optical Coherence Tomography, \textbf{Der} denotes Dermoscopy, \textbf{End} denotes Endoscopy, \textbf{Mic} denotes Microscopy Images, \textbf{US} denotes Ultrasound. Meanwhile, in each column, the best and second-best performance are marked in red and blue, respectively.}
    \begin{tabular}{l|c|c|c|c|c|c|c|c|c|c|c|c}
    \toprule[1pt]
    Model & Co & CT & DP & FP & IRI & MR & OCT & Der & End & Mic & X-Ray & US \\
    \hline
    MiniGPT-4 \cite{zhu2023minigpt} & - & 29.46 & - & 29.36 & - & 12.63 & 3- & 25.47 & - & 31.81 & 34.10 & 27.20  \\
    BLIP-2 \cite{li2023blip} & - & \textcolor{blue}{38.87} & - & 28.83 & - & 20.64 & 18.70 & 19.90 & - & \textcolor{red}{49.91} & \textcolor{blue}{48.14} & \textcolor{red}{81.79}  \\
    InstructBLIP \cite{dai2023instructblip}   & - & 35.66 & - & \textcolor{blue}{38.85} & - & 14.86 & 51.74 & \textcolor{blue}{29.81} & - & 41.32 & 36.76 & \textcolor{blue}{58.51} \\
    mPLUG-Owl \cite{ye2023mplug}  & - & 37.00 & - & \textcolor{red}{49.00} & - & 13.10 & 41.76 & 18.94 & - & 36.73 & 28.78 & 29.14  \\
    Otter \cite{li2023otter}  & - & 32.55 & - & 25.94 & - & 10.64 & 45.98 & 22.64 & - & 24.93 & 27.99 & 20.88 \\
    LLaVA \cite{liu2023visual}  & - & 38.27 & - & 20.10 & - & 4.36 & 51.23 & 14.27 & - & 23.57 & 23.67 & 20.74  \\
    LLaMA\_Adapter\_v2 \cite{gao2023llama}  & - & 36.03 & - & 30.34 & - & 17.35 & \textcolor{blue}{54.22} & 21.81 & - & 35.23 & 37.54 & 47.51  \\
    VPGTrans \cite{zhang2023transfer}  & - & 30.30 & - & 28.83 & - & 10.89 & 31.60 & 22.43 & - & 34.15 & 34.02 & 40.89  \\
    \hline
    Med-Flamingo \cite{moor2023med} & - & 22.25 & - & 36.74 & - & 14.07 & \textcolor{red}{58.57} & \textcolor{red}{39.78} & - & \textcolor{blue}{45.95} & 38.09 & 17.42 \\
    RadFM \cite{wu2023towards}  & - & \textcolor{red}{45.47} & - & 28.22 & - & 24.70 & 37.40 & 25.79 & - & 28.72 & \textcolor{red}{54.66} & 24.78 \\
    MedVInT \cite{zhang2023pmc}  & - & 37.77 & - & 9.87 & - & \textcolor{red}{30.51} & 18.54 & 20.97 & - & 23.05 & 21.86 & 25.43 \\
    LLaVA-Med \cite{li2023llava}  & - & 36.75 & - & 23.60 & - & \textcolor{blue}{24.83} & 51.51 & 25.54 & - & 30.12 & 25.04 & 16.75  \\
    \bottomrule[1pt]
    \end{tabular}%
  \label{tab:resulmsl_open-access}%
\end{table*}%

{
    \small
    \bibliographystyle{ieeenat_fullname}
    \bibliography{main}

\begin{thebibliography}{128}
\providecommand{\natexlab}[1]{#1}
\providecommand{\url}[1]{\texttt{#1}}
\expandafter\ifx\csname urlstyle\endcsname\relax
  \providecommand{\doi}[1]{doi: #1}\else
  \providecommand{\doi}{doi: \begingroup \urlstyle{rm}\Url}\fi

\bibitem[ct_({\natexlab{a}})]{ct_2d_chest_ct_scan}
Chest ct-scan images dataset.
\newblock \url{https://tianchi.aliyun.com/dataset/93929}, {\natexlab{a}}.

\bibitem[ct_({\natexlab{b}})]{ct_2d_covid_ct_covid_ct}
Covid ct dataset.
\newblock \url{https://tianchi.aliyun.com/dataset/106604}, {\natexlab{b}}.

\bibitem[der()]{dermoscopy_isic2019}
Isic 2019 challenge.
\newblock \url{https://challenge.isic-archive.com/landing/2019/}.

\bibitem[dig({\natexlab{a}})]{digital_photography_OralCancer}
Oral cancer (lips and tongue) images.
\newblock \url{https://www.kaggle.com/datasets/shivam17299/oral-cancer-lips-and-tongue-images}, {\natexlab{a}}.

\bibitem[dig({\natexlab{b}})]{digital_photography_Oral_Diseases}
Dental condition dataset.
\newblock \url{https://www.kaggle.com/datasets/salmansajid05/oral-diseases}, {\natexlab{b}}.

\bibitem[fun({\natexlab{a}})]{fundus_photography_gamma}
Glaucoma grading based on multi-modality images.
\newblock \url{https://aistudio.baidu.com/competition/detail/119/0/task-definition}, {\natexlab{a}}.

\bibitem[fun({\natexlab{b}})]{fundus_photography_glaucoma_detection}
Glaucoma detection.
\newblock \url{https://www.kaggle.com/datasets/sshikamaru/glaucoma-detection}, {\natexlab{b}}.

\bibitem[end(2016)]{endoscopy_aida_e}
Analysis of images to detect abnormalities in endoscopy.
\newblock \url{https://aidasub-cleceliachy.grand-challenge.org/Description/}, 2016.

\bibitem[dia(2023)]{diabetic-retinopathy-arranged}
Diabetic retinopathy arranged - retina images with class labels for classification.
\newblock \url{https://tianchi.aliyun.com/dataset/93926}, 2023.

\bibitem[fun(2023)]{fundus_photography_retina_cataract_dataset}
Cataract image dataset.
\newblock \url{https://www.kaggle.com/datasets/jr2ngb/cataractdataset}, 2023.

\bibitem[his(2023{\natexlab{a}})]{histopathology_bracs}
Bright challenge: Breast tumor image classification on gigapixel histopathological images.
\newblock \url{https://research.ibm.com/haifa/Workshops/BRIGHT/}, 2023{\natexlab{a}}.

\bibitem[his(2023{\natexlab{b}})]{histopathology_cell_images_malaria}
Nlm - malaria data.
\newblock \url{https://lhncbc.nlm.nih.gov/LHC-research/LHC-projects/image-processing/malaria-datasheet.html}, 2023{\natexlab{b}}.

\bibitem[mic(2023)]{microscopy_images_blood_cell_images_dataset}
Blood cell images.
\newblock \url{https://www.kaggle.com/datasets/paultimothymooney/blood-cells}, 2023.

\bibitem[oct(2023)]{oct_Retina-OCT-C8}
Retinal oct - c8 dataset.
\newblock \url{https://www.kaggle.com/datasets/obulisainaren/retinal-oct-c8/data}, 2023.

\bibitem[x_r(2023{\natexlab{a}})]{x_ray_RUS_CHN}
X-ray hand small joint classification dataset (based on bone age scoring method rus-chn).
\newblock \url{https://aistudio.baidu.com/datasetdetail/69582/0}, 2023{\natexlab{a}}.

\bibitem[x_r(2023{\natexlab{b}})]{x_ray_covid_19_image_dataset}
Covid-19 image dataset: 3 way classification - covid-19, viral pneumonia, normal.
\newblock \url{https://tianchi.aliyun.com/dataset/93853}, 2023{\natexlab{b}}.

\bibitem[Abr{\`a}moff et~al.(2013)Abr{\`a}moff, Folk, Han, Walker, Williams, Russell, Massin, Cochener, Gain, Tang, et~al.]{abramoff2013automated}
Michael~D Abr{\`a}moff, James~C Folk, Dennis~P Han, Jonathan~D Walker, David~F Williams, Stephen~R Russell, Pascale Massin, Beatrice Cochener, Philippe Gain, Li Tang, et~al.
\newblock Automated analysis of retinal images for detection of referable diabetic retinopathy.
\newblock \emph{JAMA ophthalmology}, 131\penalty0 (3):\penalty0 351--357, 2013.

\bibitem[Alayrac et~al.(2022)Alayrac, Donahue, Luc, Miech, Barr, Hasson, Lenc, Mensch, Millican, Reynolds, et~al.]{alayrac2022flamingo}
Jean-Baptiste Alayrac, Jeff Donahue, Pauline Luc, Antoine Miech, Iain Barr, Yana Hasson, Karel Lenc, Arthur Mensch, Katherine Millican, Malcolm Reynolds, et~al.
\newblock Flamingo: a visual language model for few-shot learning.
\newblock \emph{Advances in Neural Information Processing Systems}, 35:\penalty0 23716--23736, 2022.

\bibitem[Antonelli et~al.(2022)Antonelli, Reinke, Bakas, Farahani, Kopp-Schneider, Landman, Litjens, Menze, Ronneberger, Summers, et~al.]{antonelli2022medical}
Michela Antonelli, Annika Reinke, Spyridon Bakas, Keyvan Farahani, Annette Kopp-Schneider, Bennett~A Landman, Geert Litjens, Bjoern Menze, Olaf Ronneberger, Ronald~M Summers, et~al.
\newblock The medical segmentation decathlon.
\newblock \emph{Nature communications}, 13\penalty0 (1):\penalty0 4128, 2022.

\bibitem[Aresta et~al.(2019)Aresta, Ara{\'u}jo, Kwok, Chennamsetty, Safwan, Alex, Marami, Prastawa, Chan, Donovan, et~al.]{aresta2019bach}
Guilherme Aresta, Teresa Ara{\'u}jo, Scotty Kwok, Sai~Saketh Chennamsetty, Mohammed Safwan, Varghese Alex, Bahram Marami, Marcel Prastawa, Monica Chan, Michael Donovan, et~al.
\newblock Bach: Grand challenge on breast cancer histology images.
\newblock \emph{Medical image analysis}, 56:\penalty0 122--139, 2019.

\bibitem[Asraf and Islam(2021)]{asraf2021covid19}
Amanullah Asraf and Zabirul Islam.
\newblock Covid19, pneumonia and normal chest x-ray pa dataset.
\newblock \emph{Mendeley Data}, 1, 2021.

\bibitem[Bakas et~al.(2017)Bakas, Akbari, Sotiras, Bilello, Rozycki, Kirby, Freymann, Farahani, and Davatzikos]{bakas2017advancing}
Spyridon Bakas, Hamed Akbari, Aristeidis Sotiras, Michel Bilello, Martin Rozycki, Justin~S Kirby, John~B Freymann, Keyvan Farahani, and Christos Davatzikos.
\newblock Advancing the cancer genome atlas glioma mri collections with expert segmentation labels and radiomic features.
\newblock \emph{Scientific data}, 4\penalty0 (1):\penalty0 1--13, 2017.

\bibitem[Bakas et~al.(2018)Bakas, Reyes, Jakab, Bauer, Rempfler, Crimi, Shinohara, Berger, Ha, Rozycki, et~al.]{bakas2018identifying}
Spyridon Bakas, Mauricio Reyes, Andras Jakab, Stefan Bauer, Markus Rempfler, Alessandro Crimi, Russell~Takeshi Shinohara, Christoph Berger, Sung~Min Ha, Martin Rozycki, et~al.
\newblock Identifying the best machine learning algorithms for brain tumor segmentation, progression assessment, and overall survival prediction in the brats challenge.
\newblock \emph{arXiv preprint arXiv:1811.02629}, 2018.

\bibitem[Ben~Abacha et~al.(2021)Ben~Abacha, Sarrouti, Demner-Fushman, Hasan, and M{\"u}ller]{ben2021overview}
Asma Ben~Abacha, Mourad Sarrouti, Dina Demner-Fushman, Sadid~A Hasan, and Henning M{\"u}ller.
\newblock Overview of the vqa-med task at imageclef 2021: Visual question answering and generation in the medical domain.
\newblock In \emph{Proceedings of the CLEF 2021 Conference and Labs of the Evaluation Forum-working notes}. 21-24 September 2021, 2021.

\bibitem[BenO et~al.(2017)BenO, Jones, Kumar, Risdal, Rao, Sherman, Vipul, Kan, and Ben-Or]{beno2017intel}
BenO, JL Jones, H Kumar, Meg Risdal, M Rao, Vadim Sherman, Vipul, Wendy Kan, and Yau Ben-Or.
\newblock Intel \& mobileodt cervical cancer screening.
\newblock \url{https://kaggle.com/competitions/intel-mobileodt-cervical-cancer-screening}, 2017.

\bibitem[Borkowski et~al.(2019)Borkowski, Bui, Thomas, Wilson, DeLand, and Mastorides]{borkowski2019lung}
Andrew~A Borkowski, Marilyn~M Bui, L~Brannon Thomas, Catherine~P Wilson, Lauren~A DeLand, and Stephen~M Mastorides.
\newblock Lung and colon cancer histopathological image dataset (lc25000).
\newblock \emph{arXiv preprint arXiv:1912.12142}, 2019.

\bibitem[Byeon et~al.(2022)Byeon, Park, Kim, Lee, Baek, and Kim]{kakaobrain2022coyo-700m}
Minwoo Byeon, Beomhee Park, Haecheon Kim, Sungjun Lee, Woonhyuk Baek, and Saehoon Kim.
\newblock Coyo-700m: Image-text pair dataset.
\newblock \url{https://github.com/kakaobrain/coyo-dataset}, 2022.

\bibitem[Carass et~al.(2017)Carass, Roy, Jog, Cuzzocreo, Magrath, Gherman, Button, Nguyen, Prados, Sudre, et~al.]{carass2017longitudinal}
Aaron Carass, Snehashis Roy, Amod Jog, Jennifer~L Cuzzocreo, Elizabeth Magrath, Adrian Gherman, Julia Button, James Nguyen, Ferran Prados, Carole~H Sudre, et~al.
\newblock Longitudinal multiple sclerosis lesion segmentation: resource and challenge.
\newblock \emph{NeuroImage}, 148:\penalty0 77--102, 2017.

\bibitem[Carole~Sudre(2021)]{VALDO_Task2}
Kimberlin van~Wijnen Carole~Sudre.
\newblock Where is valdo - vascular lesions detection challenge 2021.
\newblock \url{https://valdo.grand-challenge.org/}, 2021.

\bibitem[Cen et~al.(2021)Cen, Ji, Lin, Ju, Lin, Li, Wang, Yang, Liu, Tan, et~al.]{cen2021automatic}
Ling-Ping Cen, Jie Ji, Jian-Wei Lin, Si-Tong Ju, Hong-Jie Lin, Tai-Ping Li, Yun Wang, Jian-Feng Yang, Yu-Fen Liu, Shaoying Tan, et~al.
\newblock Automatic detection of 39 fundus diseases and conditions in retinal photographs using deep neural networks.
\newblock \emph{Nature communications}, 12\penalty0 (1):\penalty0 4828, 2021.

\bibitem[Cereda et~al.(2016)Cereda, Christensen, Campbell, Mishra, Mlynash, Levi, Straka, Wintermark, Bammer, Albers, et~al.]{cereda2016benchmarking}
Carlo~W Cereda, S{\o}ren Christensen, Bruce~CV Campbell, Nishant~K Mishra, Michael Mlynash, Christopher Levi, Matus Straka, Max Wintermark, Roland Bammer, Gregory~W Albers, et~al.
\newblock A benchmarking tool to evaluate computer tomography perfusion infarct core predictions against a dwi standard.
\newblock \emph{Journal of Cerebral Blood Flow \& Metabolism}, 36\penalty0 (10):\penalty0 1780--1789, 2016.

\bibitem[Changpinyo et~al.(2021)Changpinyo, Sharma, Ding, and Soricut]{changpinyo2021conceptual}
Soravit Changpinyo, Piyush Sharma, Nan Ding, and Radu Soricut.
\newblock Conceptual 12m: Pushing web-scale image-text pre-training to recognize long-tail visual concepts.
\newblock In \emph{Proceedings of the IEEE/CVF Conference on Computer Vision and Pattern Recognition}, pages 3558--3568, 2021.

\bibitem[Chen(2018)]{chen2018knee}
Pingjun Chen.
\newblock Knee osteoarthritis severity grading dataset.
\newblock \emph{Mendeley Data}, 1:\penalty0 21--23, 2018.

\bibitem[Chen et~al.(2015)Chen, Fang, Lin, Vedantam, Gupta, Doll{\'a}r, and Zitnick]{chen2015microsoft}
Xinlei Chen, Hao Fang, Tsung-Yi Lin, Ramakrishna Vedantam, Saurabh Gupta, Piotr Doll{\'a}r, and C~Lawrence Zitnick.
\newblock Microsoft coco captions: Data collection and evaluation server.
\newblock \emph{arXiv preprint arXiv:1504.00325}, 2015.

\bibitem[Chowdhury et~al.(2020)Chowdhury, Rahman, Khandakar, Mazhar, Kadir, Mahbub, Islam, Khan, Iqbal, Al~Emadi, et~al.]{chowdhury2020can}
Muhammad~EH Chowdhury, Tawsifur Rahman, Amith Khandakar, Rashid Mazhar, Muhammad~Abdul Kadir, Zaid~Bin Mahbub, Khandakar~Reajul Islam, Muhammad~Salman Khan, Atif Iqbal, Nasser Al~Emadi, et~al.
\newblock Can ai help in screening viral and covid-19 pneumonia?
\newblock \emph{Ieee Access}, 8:\penalty0 132665--132676, 2020.

\bibitem[Codella et~al.(2019)Codella, Rotemberg, Tschandl, Celebi, Dusza, Gutman, Helba, Kalloo, Liopyris, Marchetti, et~al.]{codella2019skin}
Noel Codella, Veronica Rotemberg, Philipp Tschandl, M~Emre Celebi, Stephen Dusza, David Gutman, Brian Helba, Aadi Kalloo, Konstantinos Liopyris, Michael Marchetti, et~al.
\newblock Skin lesion analysis toward melanoma detection 2018: A challenge hosted by the international skin imaging collaboration (isic).
\newblock \emph{arXiv preprint arXiv:1902.03368}, 2019.

\bibitem[Cohen et~al.(2020)Cohen, Morrison, Dao, Roth, Duong, and Ghassemi]{cohen2020covid}
Joseph~Paul Cohen, Paul Morrison, Lan Dao, Karsten Roth, Tim~Q Duong, and Marzyeh Ghassemi.
\newblock Covid-19 image data collection: Prospective predictions are the future.
\newblock \emph{arXiv preprint arXiv:2006.11988}, 2020.

\bibitem[Commowick et~al.(2021)Commowick, Kain, Casey, Ameli, Ferr{\'e}, Kerbrat, Tourdias, Cervenansky, Camarasu-Pop, Glatard, et~al.]{commowick2021multiple}
Olivier Commowick, Micha{\"e}l Kain, Romain Casey, Roxana Ameli, Jean-Christophe Ferr{\'e}, Anne Kerbrat, Thomas Tourdias, Fr{\'e}d{\'e}ric Cervenansky, Sorina Camarasu-Pop, Tristan Glatard, et~al.
\newblock Multiple sclerosis lesions segmentation from multiple experts: The miccai 2016 challenge dataset.
\newblock \emph{Neuroimage}, 244:\penalty0 118589, 2021.

\bibitem[Cukierski(2018)]{cukierski2018histopathologic}
Will Cukierski.
\newblock Histopathologic cancer detection.
\newblock \url{https://kaggle.com/competitions/histopathologic-cancer-detection}, 2018.

\bibitem[Da et~al.(2022)Da, Huang, Li, Zuo, Zhang, Liu, Chen, Li, Xu, Hu, et~al.]{da2022digestpath}
Qian Da, Xiaodi Huang, Zhongyu Li, Yanfei Zuo, Chenbin Zhang, Jingxin Liu, Wen Chen, Jiahui Li, Dou Xu, Zhiqiang Hu, et~al.
\newblock Digestpath: A benchmark dataset with challenge review for the pathological detection and segmentation of digestive-system.
\newblock \emph{Medical Image Analysis}, 80:\penalty0 102485, 2022.

\bibitem[Dai et~al.(2023)Dai, Li, Li, Tiong, Zhao, Wang, Li, Fung, and Hoi]{dai2023instructblip}
Wenliang Dai, Junnan Li, Dongxu Li, Anthony Meng~Huat Tiong, Junqi Zhao, Weisheng Wang, Boyang Li, Pascale Fung, and Steven Hoi.
\newblock Instructblip: Towards general-purpose vision-language models with instruction tuning.
\newblock \emph{arXiv preprint arXiv:2305.06500}, 2023.

\bibitem[De~Vente et~al.(2023)De~Vente, Vermeer, Jaccard, Wang, Sun, Khader, Truhn, Aimyshev, Zhanibekuly, Le, et~al.]{de2023airogs}
Coen De~Vente, Koenraad~A Vermeer, Nicolas Jaccard, He Wang, Hongyi Sun, Firas Khader, Daniel Truhn, Temirgali Aimyshev, Yerkebulan Zhanibekuly, Tien-Dung Le, et~al.
\newblock Airogs: Artificial intelligence for robust glaucoma screening challenge.
\newblock \emph{IEEE Transactions on Medical Imaging}, 2023.

\bibitem[Decenci{\`e}re et~al.(2014)Decenci{\`e}re, Zhang, Cazuguel, Lay, Cochener, Trone, Gain, Ordonez, Massin, Erginay, et~al.]{decenciere2014feedback}
Etienne Decenci{\`e}re, Xiwei Zhang, Guy Cazuguel, Bruno Lay, B{\'e}atrice Cochener, Caroline Trone, Philippe Gain, Richard Ordonez, Pascale Massin, Ali Erginay, et~al.
\newblock Feedback on a publicly distributed image database: the messidor database.
\newblock \emph{Image Analysis \& Stereology}, 33\penalty0 (3):\penalty0 231--234, 2014.

\bibitem[Diaz-Pinto et~al.(2019)Diaz-Pinto, Morales, Naranjo, K{\"o}hler, Mossi, and Navea]{diaz2019cnns}
Andres Diaz-Pinto, Sandra Morales, Valery Naranjo, Thomas K{\"o}hler, Jose~M Mossi, and Amparo Navea.
\newblock Cnns for automatic glaucoma assessment using fundus images: an extensive validation.
\newblock \emph{Biomedical engineering online}, 18:\penalty0 1--19, 2019.

\bibitem[Fang et~al.(2022)Fang, Li, Fu, Sun, Cao, Lin, Son, Kim, Quellec, Matta, et~al.]{fang2022adam}
Huihui Fang, Fei Li, Huazhu Fu, Xu Sun, Xingxing Cao, Fengbin Lin, Jaemin Son, Sunho Kim, Gwenole Quellec, Sarah Matta, et~al.
\newblock Adam challenge: Detecting age-related macular degeneration from fundus images.
\newblock \emph{IEEE Transactions on Medical Imaging}, 41\penalty0 (10):\penalty0 2828--2847, 2022.

\bibitem[Fang et~al.(2023)Fang, Li, Wu, Fu, Sun, Orlando, Bogunovi{\'c}, Zhang, and Xu]{fang2023palm}
Huihui Fang, Fei Li, Junde Wu, Huazhu Fu, Xu Sun, Jos{\'e}~Ignacio Orlando, Hrvoje Bogunovi{\'c}, Xiulan Zhang, and Yanwu Xu.
\newblock Palm: Open fundus photograph dataset with pathologic myopia recognition and anatomical structure annotation.
\newblock \emph{arXiv preprint arXiv:2305.07816}, 2023.

\bibitem[Fusek(2018)]{fusek2018pupil}
Radovan Fusek.
\newblock Pupil localization using geodesic distance.
\newblock In \emph{Advances in Visual Computing: 13th International Symposium, ISVC 2018, Las Vegas, NV, USA, November 19--21, 2018, Proceedings 13}, pages 433--444. Springer, 2018.

\bibitem[Gao et~al.(2023)Gao, Han, Zhang, Lin, Geng, Zhou, Zhang, Lu, He, Yue, et~al.]{gao2023llama}
Peng Gao, Jiaming Han, Renrui Zhang, Ziyi Lin, Shijie Geng, Aojun Zhou, Wei Zhang, Pan Lu, Conghui He, Xiangyu Yue, et~al.
\newblock Llama-adapter v2: Parameter-efficient visual instruction model.
\newblock \emph{arXiv preprint arXiv:2304.15010}, 2023.

\bibitem[Giotis et~al.(2015)Giotis, Molders, Land, Biehl, Jonkman, and Petkov]{giotis2015med}
Ioannis Giotis, Nynke Molders, Sander Land, Michael Biehl, Marcel~F Jonkman, and Nicolai Petkov.
\newblock Med-node: A computer-assisted melanoma diagnosis system using non-dermoscopic images.
\newblock \emph{Expert systems with applications}, 42\penalty0 (19):\penalty0 6578--6585, 2015.

\bibitem[Groh et~al.(2021)Groh, Harris, Soenksen, Lau, Han, Kim, Koochek, and Badri]{groh2021evaluating}
Matthew Groh, Caleb Harris, Luis Soenksen, Felix Lau, Rachel Han, Aerin Kim, Arash Koochek, and Omar Badri.
\newblock Evaluating deep neural networks trained on clinical images in dermatology with the fitzpatrick 17k dataset.
\newblock In \emph{Proceedings of the IEEE/CVF Conference on Computer Vision and Pattern Recognition}, pages 1820--1828, 2021.

\bibitem[Groh et~al.(2022)Groh, Harris, Daneshjou, Badri, and Koochek]{groh2022towards}
Matthew Groh, Caleb Harris, Roxana Daneshjou, Omar Badri, and Arash Koochek.
\newblock Towards transparency in dermatology image datasets with skin tone annotations by experts, crowds, and an algorithm.
\newblock \emph{arXiv preprint arXiv:2207.02942}, 2022.

\bibitem[Gupta and Gupta(2019)]{gupta2019isbi}
Anubha Gupta and Ritu Gupta.
\newblock Isbi 2019 c-nmc challenge: Classification in cancer cell imaging.
\newblock \emph{Select Proceedings}, 2019.

\bibitem[Gutman et~al.(2016)Gutman, Codella, Celebi, Helba, Marchetti, Mishra, and Halpern]{gutman2016skin}
David Gutman, Noel~CF Codella, Emre Celebi, Brian Helba, Michael Marchetti, Nabin Mishra, and Allan Halpern.
\newblock Skin lesion analysis toward melanoma detection: A challenge at the international symposium on biomedical imaging (isbi) 2016, hosted by the international skin imaging collaboration (isic).
\newblock \emph{arXiv preprint arXiv:1605.01397}, 2016.

\bibitem[Hakim et~al.(2021)Hakim, Christensen, Winzeck, Lansberg, Parsons, Lucas, Robben, Wiest, Reyes, and Zaharchuk]{hakim2021predicting}
Arsany Hakim, S{\o}ren Christensen, Stefan Winzeck, Maarten~G Lansberg, Mark~W Parsons, Christian Lucas, David Robben, Roland Wiest, Mauricio Reyes, and Greg Zaharchuk.
\newblock Predicting infarct core from computed tomography perfusion in acute ischemia with machine learning: Lessons from the isles challenge.
\newblock \emph{Stroke}, 52\penalty0 (7):\penalty0 2328--2337, 2021.

\bibitem[Hamada(2020)]{hamada2020br35h}
Ahmed Hamada.
\newblock Br35h:: Brain tumor detection 2020.
\newblock \emph{Version 5}, 2020.

\bibitem[Harrar(2014)]{harrar2014texture}
Khaled Harrar.
\newblock Texture characterization of bone radiograph images. application to osteoporosis diagnosis.
\newblock 2014.

\bibitem[Hernandez~Petzsche et~al.(2022)Hernandez~Petzsche, de~la Rosa, Hanning, Wiest, Valenzuela, Reyes, Meyer, Liew, Kofler, Ezhov, et~al.]{hernandez2022isles}
Moritz~R Hernandez~Petzsche, Ezequiel de~la Rosa, Uta Hanning, Roland Wiest, Waldo Valenzuela, Mauricio Reyes, Maria Meyer, Sook-Lei Liew, Florian Kofler, Ivan Ezhov, et~al.
\newblock Isles 2022: A multi-center magnetic resonance imaging stroke lesion segmentation dataset.
\newblock \emph{Scientific data}, 9\penalty0 (1):\penalty0 762, 2022.

\bibitem[Islam et~al.(2022)Islam, Hussain, Chowdhury, and Riazul~Islam]{islam2022web}
Towhidul Islam, Mohammad~Arafat Hussain, Forhad Uddin~Hasan Chowdhury, and BM Riazul~Islam.
\newblock A web-scraped skin image database of monkeypox, chickenpox, smallpox, cowpox, and measles.
\newblock \emph{biorxiv}, pages 2022--08, 2022.

\bibitem[Jaeger et~al.(2014)Jaeger, Candemir, Antani, W{\'a}ng, Lu, and Thoma]{jaeger2014two}
Stefan Jaeger, Sema Candemir, Sameer Antani, Y{\`\i}-Xi{\'a}ng~J W{\'a}ng, Pu-Xuan Lu, and George Thoma.
\newblock Two public chest x-ray datasets for computer-aided screening of pulmonary diseases.
\newblock \emph{Quantitative imaging in medicine and surgery}, 4\penalty0 (6):\penalty0 475, 2014.

\bibitem[Javadi and Mirroshandel(2019)]{javadi2019novel}
Soroush Javadi and Seyed~Abolghasem Mirroshandel.
\newblock A novel deep learning method for automatic assessment of human sperm images.
\newblock \emph{Computers in biology and medicine}, 109:\penalty0 182--194, 2019.

\bibitem[Ji et~al.(2022)Ji, Bai, Ge, Yang, Zhu, Zhang, Li, Zhanng, Ma, Wan, et~al.]{ji2022amos}
Yuanfeng Ji, Haotian Bai, Chongjian Ge, Jie Yang, Ye Zhu, Ruimao Zhang, Zhen Li, Lingyan Zhanng, Wanling Ma, Xiang Wan, et~al.
\newblock Amos: A large-scale abdominal multi-organ benchmark for versatile medical image segmentation.
\newblock \emph{Advances in Neural Information Processing Systems}, 35:\penalty0 36722--36732, 2022.

\bibitem[Karthik et~al.(2019)Karthik, Maggie, and Dane]{karthik_2019}
Karthik, Maggie, and Sohier Dane.
\newblock Aptos 2019 blindness detection.
\newblock Kaggle, 2019.

\bibitem[Kather et~al.(2018)Kather, Halama, and Marx]{kather2018100}
Jakob~Nikolas Kather, Niels Halama, and Alexander Marx.
\newblock 100,000 histological images of human colorectal cancer and healthy tissue.
\newblock \emph{Zenodo10}, 5281, 2018.

\bibitem[Kermany et~al.(2018)Kermany, Goldbaum, Cai, Valentim, Liang, Baxter, McKeown, Yang, Wu, Yan, et~al.]{kermany2018identifying}
Daniel~S Kermany, Michael Goldbaum, Wenjia Cai, Carolina~CS Valentim, Huiying Liang, Sally~L Baxter, Alex McKeown, Ge Yang, Xiaokang Wu, Fangbing Yan, et~al.
\newblock Identifying medical diagnoses and treatable diseases by image-based deep learning.
\newblock \emph{cell}, 172\penalty0 (5):\penalty0 1122--1131, 2018.

\bibitem[Kistler et~al.(2013)Kistler, Bonaretti, Pfahrer, Niklaus, and B{\"u}chler]{kistler2013virtual}
Michael Kistler, Serena Bonaretti, Marcel Pfahrer, Roman Niklaus, and Philippe B{\"u}chler.
\newblock The virtual skeleton database: an open access repository for biomedical research and collaboration.
\newblock \emph{Journal of medical Internet research}, 15\penalty0 (11):\penalty0 e245, 2013.

\bibitem[Krishna et~al.(2017)Krishna, Zhu, Groth, Johnson, Hata, Kravitz, Chen, Kalantidis, Li, Shamma, et~al.]{krishna2017visual}
Ranjay Krishna, Yuke Zhu, Oliver Groth, Justin Johnson, Kenji Hata, Joshua Kravitz, Stephanie Chen, Yannis Kalantidis, Li-Jia Li, David~A Shamma, et~al.
\newblock Visual genome: Connecting language and vision using crowdsourced dense image annotations.
\newblock \emph{International journal of computer vision}, 123:\penalty0 32--73, 2017.

\bibitem[Kuckreja et~al.(2023)Kuckreja, Danish, Naseer, Das, Khan, and Khan]{kuckreja2023geochat}
Kartik Kuckreja, Muhammad~Sohail Danish, Muzammal Naseer, Abhijit Das, Salman Khan, and Fahad~Shahbaz Khan.
\newblock Geochat: Grounded large vision-language model for remote sensing.
\newblock \emph{arXiv preprint arXiv:2311.15826}, 2023.

\bibitem[Kuijf and Bennink(2019)]{kuijf2019grand}
Hugo~J Kuijf and E Bennink.
\newblock Grand challenge on mr brain segmentation at miccai 2018, 2019.

\bibitem[Lau et~al.(2018)Lau, Gayen, Ben~Abacha, and Demner-Fushman]{lau2018dataset}
Jason~J Lau, Soumya Gayen, Asma Ben~Abacha, and Dina Demner-Fushman.
\newblock A dataset of clinically generated visual questions and answers about radiology images.
\newblock \emph{Scientific data}, 5\penalty0 (1):\penalty0 1--10, 2018.

\bibitem[Li et~al.(2023{\natexlab{a}})Li, Zhang, Chen, Wang, Yang, and Liu]{li2023otter}
Bo Li, Yuanhan Zhang, Liangyu Chen, Jinghao Wang, Jingkang Yang, and Ziwei Liu.
\newblock Otter: A multi-modal model with in-context instruction tuning.
\newblock \emph{arXiv preprint arXiv:2305.03726}, 2023{\natexlab{a}}.

\bibitem[Li et~al.(2023{\natexlab{b}})Li, Wong, Zhang, Usuyama, Liu, Yang, Naumann, Poon, and Gao]{li2023llava}
Chunyuan Li, Cliff Wong, Sheng Zhang, Naoto Usuyama, Haotian Liu, Jianwei Yang, Tristan Naumann, Hoifung Poon, and Jianfeng Gao.
\newblock Llava-med: Training a large language-and-vision assistant for biomedicine in one day.
\newblock \emph{arXiv preprint arXiv:2306.00890}, 2023{\natexlab{b}}.

\bibitem[Li et~al.(2020)Li, Song, Chen, Xiong, Li, Zhong, Tang, Fan, Lam, Pan, et~al.]{li2020development}
Fei Li, Diping Song, Han Chen, Jian Xiong, Xingyi Li, Hua Zhong, Guangxian Tang, Sujie Fan, Dennis~SC Lam, Weihua Pan, et~al.
\newblock Development and clinical deployment of a smartphone-based visual field deep learning system for glaucoma detection.
\newblock \emph{NPJ digital medicine}, 3\penalty0 (1):\penalty0 123, 2020.

\bibitem[Li et~al.(2023{\natexlab{c}})Li, Li, Savarese, and Hoi]{li2023blip}
Junnan Li, Dongxu Li, Silvio Savarese, and Steven Hoi.
\newblock Blip-2: Bootstrapping language-image pre-training with frozen image encoders and large language models.
\newblock \emph{arXiv preprint arXiv:2301.12597}, 2023{\natexlab{c}}.

\bibitem[Lin et~al.(2023)Lin, Zhao, Zhang, Wu, Zhang, Wang, and Xie]{lin2023pmc}
Weixiong Lin, Ziheng Zhao, Xiaoman Zhang, Chaoyi Wu, Ya Zhang, Yanfeng Wang, and Weidi Xie.
\newblock Pmc-clip: Contrastive language-image pre-training using biomedical documents.
\newblock \emph{arXiv preprint arXiv:2303.07240}, 2023.

\bibitem[Liu et~al.(2021)Liu, Zhan, Xu, Ma, Yang, and Wu]{liu2021slake}
Bo Liu, Li-Ming Zhan, Li Xu, Lin Ma, Yan Yang, and Xiao-Ming Wu.
\newblock Slake: A semantically-labeled knowledge-enhanced dataset for medical visual question answering.
\newblock In \emph{2021 IEEE 18th International Symposium on Biomedical Imaging (ISBI)}, pages 1650--1654. IEEE, 2021.

\bibitem[Liu et~al.(2019)Liu, Han, Li, Ha, Peng, Meng, and He]{liu2019self}
Chi Liu, Xiaotong Han, Zhixi Li, Jason Ha, Guankai Peng, Wei Meng, and Mingguang He.
\newblock A self-adaptive deep learning method for automated eye laterality detection based on color fundus photography.
\newblock \emph{Plos one}, 14\penalty0 (9):\penalty0 e0222025, 2019.

\bibitem[Liu et~al.(2023)Liu, Li, Wu, and Lee]{liu2023visual}
Haotian Liu, Chunyuan Li, Qingyang Wu, and Yong~Jae Lee.
\newblock Visual instruction tuning.
\newblock \emph{arXiv preprint arXiv:2304.08485}, 2023.

\bibitem[Liu et~al.(2022)Liu, Wang, Wu, Dai, Fang, Yan, Son, Tang, Li, Gao, et~al.]{liu2022deepdrid}
Ruhan Liu, Xiangning Wang, Qiang Wu, Ling Dai, Xi Fang, Tao Yan, Jaemin Son, Shiqi Tang, Jiang Li, Zijian Gao, et~al.
\newblock Deepdrid: Diabetic retinopathy—grading and image quality estimation challenge.
\newblock \emph{Patterns}, 3\penalty0 (6), 2022.

\bibitem[Maier et~al.(2017)Maier, Menze, Von~der Gablentz, H{\"a}ni, Heinrich, Liebrand, Winzeck, Basit, Bentley, Chen, et~al.]{maier2017isles}
Oskar Maier, Bjoern~H Menze, Janina Von~der Gablentz, Levin H{\"a}ni, Mattias~P Heinrich, Matthias Liebrand, Stefan Winzeck, Abdul Basit, Paul Bentley, Liang Chen, et~al.
\newblock Isles 2015-a public evaluation benchmark for ischemic stroke lesion segmentation from multispectral mri.
\newblock \emph{Medical image analysis}, 35:\penalty0 250--269, 2017.

\bibitem[Mei et~al.(2022)Mei, Liu, Robson, Marinelli, Huang, Doshi, Jacobi, Cao, Link, Yang, et~al.]{mei2022radimagenet}
Xueyan Mei, Zelong Liu, Philip~M Robson, Brett Marinelli, Mingqian Huang, Amish Doshi, Adam Jacobi, Chendi Cao, Katherine~E Link, Thomas Yang, et~al.
\newblock Radimagenet: an open radiologic deep learning research dataset for effective transfer learning.
\newblock \emph{Radiology: Artificial Intelligence}, 4\penalty0 (5):\penalty0 e210315, 2022.

\bibitem[Mendon{\c{c}}a et~al.(2013)Mendon{\c{c}}a, Ferreira, Marques, Marcal, and Rozeira]{mendoncca2013ph}
Teresa Mendon{\c{c}}a, Pedro~M Ferreira, Jorge~S Marques, Andr{\'e}~RS Marcal, and Jorge Rozeira.
\newblock Ph$^{2}$-a dermoscopic image database for research and benchmarking.
\newblock In \emph{2013 35th annual international conference of the IEEE engineering in medicine and biology society (EMBC)}, pages 5437--5440. IEEE, 2013.

\bibitem[Mendrik et~al.(2015)Mendrik, Vincken, Kuijf, Breeuwer, Bouvy, De~Bresser, Alansary, De~Bruijne, Carass, El-Baz, et~al.]{mendrik2015mrbrains}
Adri{\"e}nne~M Mendrik, Koen~L Vincken, Hugo~J Kuijf, Marcel Breeuwer, Willem~H Bouvy, Jeroen De~Bresser, Amir Alansary, Marleen De~Bruijne, Aaron Carass, Ayman El-Baz, et~al.
\newblock Mrbrains challenge: online evaluation framework for brain image segmentation in 3t mri scans.
\newblock \emph{Computational intelligence and neuroscience}, 2015:\penalty0 1--1, 2015.

\bibitem[Menze et~al.(2014)Menze, Jakab, Bauer, Kalpathy-Cramer, Farahani, Kirby, Burren, Porz, Slotboom, Wiest, et~al.]{menze2014multimodal}
Bjoern~H Menze, Andras Jakab, Stefan Bauer, Jayashree Kalpathy-Cramer, Keyvan Farahani, Justin Kirby, Yuliya Burren, Nicole Porz, Johannes Slotboom, Roland Wiest, et~al.
\newblock The multimodal brain tumor image segmentation benchmark (brats).
\newblock \emph{IEEE transactions on medical imaging}, 34\penalty0 (10):\penalty0 1993--2024, 2014.

\bibitem[Moor et~al.(2023)Moor, Huang, Wu, Yasunaga, Zakka, Dalmia, Reis, Rajpurkar, and Leskovec]{moor2023med}
Michael Moor, Qian Huang, Shirley Wu, Michihiro Yasunaga, Cyril Zakka, Yash Dalmia, Eduardo~Pontes Reis, Pranav Rajpurkar, and Jure Leskovec.
\newblock Med-flamingo: a multimodal medical few-shot learner.
\newblock \emph{arXiv preprint arXiv:2307.15189}, 2023.

\bibitem[Mu et~al.(2024)Mu, Zhang, Hu, Wang, Ding, Jin, Wang, Dai, Qiao, and Luo]{mu2024embodiedgpt}
Yao Mu, Qinglong Zhang, Mengkang Hu, Wenhai Wang, Mingyu Ding, Jun Jin, Bin Wang, Jifeng Dai, Yu Qiao, and Ping Luo.
\newblock Embodiedgpt: Vision-language pre-training via embodied chain of thought.
\newblock \emph{Advances in Neural Information Processing Systems}, 36, 2024.

\bibitem[Nanni et~al.(2016)Nanni, Paci, Caetano~dos Santos, Skottman, Juuti-Uusitalo, and Hyttinen]{nanni2016texture}
Loris Nanni, Michelangelo Paci, Florentino~Luciano Caetano~dos Santos, Heli Skottman, Kati Juuti-Uusitalo, and Jari Hyttinen.
\newblock Texture descriptors ensembles enable image-based classification of maturation of human stem cell-derived retinal pigmented epithelium.
\newblock \emph{PLoS One}, 11\penalty0 (2):\penalty0 e0149399, 2016.

\bibitem[Narendran()]{AVNickingQuantification}
Thivya Narendran.
\newblock Image set for retinal artery-vein nicking assessment.
\newblock \url{https://people.eng.unimelb.edu.au/thivun/projects/AV_nicking_quantification/}.

\bibitem[Orlando et~al.(2020)Orlando, Fu, Breda, Van~Keer, Bathula, Diaz-Pinto, Fang, Heng, Kim, Lee, et~al.]{orlando2020refuge}
Jos{\'e}~Ignacio Orlando, Huazhu Fu, Jo{\~a}o~Barbosa Breda, Karel Van~Keer, Deepti~R Bathula, Andr{\'e}s Diaz-Pinto, Ruogu Fang, Pheng-Ann Heng, Jeyoung Kim, JoonHo Lee, et~al.
\newblock Refuge challenge: A unified framework for evaluating automated methods for glaucoma assessment from fundus photographs.
\newblock \emph{Medical image analysis}, 59:\penalty0 101570, 2020.

\bibitem[Orlov et~al.(2010)Orlov, Chen, Eckley, Macura, Shamir, Jaffe, and Goldberg]{orlov2010automatic}
Nikita~V Orlov, Wayne~W Chen, David~Mark Eckley, Tomasz~J Macura, Lior Shamir, Elaine~S Jaffe, and Ilya~G Goldberg.
\newblock Automatic classification of lymphoma images with transform-based global features.
\newblock \emph{IEEE Transactions on Information Technology in Biomedicine}, 14\penalty0 (4):\penalty0 1003--1013, 2010.

\bibitem[Ouyang et~al.(2022)Ouyang, Wu, Jiang, Almeida, Wainwright, Mishkin, Zhang, Agarwal, Slama, Ray, et~al.]{ouyang2022training}
Long Ouyang, Jeffrey Wu, Xu Jiang, Diogo Almeida, Carroll Wainwright, Pamela Mishkin, Chong Zhang, Sandhini Agarwal, Katarina Slama, Alex Ray, et~al.
\newblock Training language models to follow instructions with human feedback.
\newblock \emph{Advances in Neural Information Processing Systems}, 35:\penalty0 27730--27744, 2022.

\bibitem[Pacheco and Krohling(2020)]{pacheco2020impact}
Andre~GC Pacheco and Renato~A Krohling.
\newblock The impact of patient clinical information on automated skin cancer detection.
\newblock \emph{Computers in biology and medicine}, 116:\penalty0 103545, 2020.

\bibitem[Peng et~al.(2023)Peng, Li, He, Galley, and Gao]{peng2023instruction}
Baolin Peng, Chunyuan Li, Pengcheng He, Michel Galley, and Jianfeng Gao.
\newblock Instruction tuning with gpt-4.
\newblock \emph{arXiv preprint arXiv:2304.03277}, 2023.

\bibitem[Phoulady and Mouton(2018)]{phoulady2018new}
Hady~Ahmady Phoulady and Peter~R. Mouton.
\newblock A new cervical cytology dataset for nucleus detection and image classification (cervix93) and methods for cervical nucleus detection, 2018.

\bibitem[Pogorelov et~al.(2017)Pogorelov, Randel, Griwodz, Eskeland, de~Lange, Johansen, Spampinato, Dang-Nguyen, Lux, Schmidt, et~al.]{pogorelov2017kvasir}
Konstantin Pogorelov, Kristin~Ranheim Randel, Carsten Griwodz, Sigrun~Losada Eskeland, Thomas de Lange, Dag Johansen, Concetto Spampinato, Duc-Tien Dang-Nguyen, Mathias Lux, Peter~Thelin Schmidt, et~al.
\newblock Kvasir: A multi-class image dataset for computer aided gastrointestinal disease detection.
\newblock In \emph{Proceedings of the 8th ACM on Multimedia Systems Conference}, pages 164--169, 2017.

\bibitem[Prabhushankar et~al.(2022)Prabhushankar, Kokilepersaud, Logan, Trejo~Corona, AlRegib, and Wykoff]{prabhushankar2022olives}
Mohit Prabhushankar, Kiran Kokilepersaud, Yash-yee Logan, Stephanie Trejo~Corona, Ghassan AlRegib, and Charles Wykoff.
\newblock Olives dataset: Ophthalmic labels for investigating visual eye semantics.
\newblock \emph{Advances in Neural Information Processing Systems}, 35:\penalty0 9201--9216, 2022.

\bibitem[Rajpurkar et~al.(2017)Rajpurkar, Irvin, Bagul, Ding, Duan, Mehta, Yang, Zhu, Laird, Ball, et~al.]{rajpurkar2017mura}
Pranav Rajpurkar, Jeremy Irvin, Aarti Bagul, Daisy Ding, Tony Duan, Hershel Mehta, Brandon Yang, Kaylie Zhu, Dillon Laird, Robyn~L Ball, et~al.
\newblock Mura: Large dataset for abnormality detection in musculoskeletal radiographs.
\newblock \emph{arXiv preprint arXiv:1712.06957}, 2017.

\bibitem[Rotemberg et~al.(2021)Rotemberg, Kurtansky, Betz-Stablein, Caffery, Chousakos, Codella, Combalia, Dusza, Guitera, Gutman, et~al.]{rotemberg2021patient}
Veronica Rotemberg, Nicholas Kurtansky, Brigid Betz-Stablein, Liam Caffery, Emmanouil Chousakos, Noel Codella, Marc Combalia, Stephen Dusza, Pascale Guitera, David Gutman, et~al.
\newblock A patient-centric dataset of images and metadata for identifying melanomas using clinical context.
\newblock \emph{Scientific data}, 8\penalty0 (1):\penalty0 34, 2021.

\bibitem[Saha et~al.(2021)Saha, Hosseinzadeh, and Huisman]{saha2021end}
Anindo Saha, Matin Hosseinzadeh, and Henkjan Huisman.
\newblock End-to-end prostate cancer detection in bpmri via 3d cnns: effects of attention mechanisms, clinical priori and decoupled false positive reduction.
\newblock \emph{Medical image analysis}, 73:\penalty0 102155, 2021.

\bibitem[Scarpa et~al.(2008)Scarpa, Grisan, and Ruggeri]{scarpa2008automatic}
Fabio Scarpa, Enrico Grisan, and Alfredo Ruggeri.
\newblock Automatic recognition of corneal nerve structures in images from confocal microscopy.
\newblock \emph{Investigative ophthalmology \& visual science}, 49\penalty0 (11):\penalty0 4801--4807, 2008.

\bibitem[Scarpa et~al.(2011)Scarpa, Zheng, Ohashi, and Ruggeri]{scarpa2011automatic}
Fabio Scarpa, Xiaodong Zheng, Yuichi Ohashi, and Alfredo Ruggeri.
\newblock Automatic evaluation of corneal nerve tortuosity in images from in vivo confocal microscopy.
\newblock \emph{Investigative ophthalmology \& visual science}, 52\penalty0 (9):\penalty0 6404--6408, 2011.

\bibitem[Schuhmann et~al.(2021)Schuhmann, Vencu, Beaumont, Kaczmarczyk, Mullis, Katta, Coombes, Jitsev, and Komatsuzaki]{schuhmann2021laion}
Christoph Schuhmann, Richard Vencu, Romain Beaumont, Robert Kaczmarczyk, Clayton Mullis, Aarush Katta, Theo Coombes, Jenia Jitsev, and Aran Komatsuzaki.
\newblock Laion-400m: Open dataset of clip-filtered 400 million image-text pairs.
\newblock \emph{arXiv preprint arXiv:2111.02114}, 2021.

\bibitem[{\c{S}}evik et~al.(2014){\c{S}}evik, K{\"o}se, Berber, and Erd{\"o}l]{fundus_photography_drimdb}
U{\u{g}}ur {\c{S}}evik, Cemal K{\"o}se, Tolga Berber, and Hidayet Erd{\"o}l.
\newblock Identification of suitable fundus images using automated quality assessment methods.
\newblock \emph{Journal of biomedical optics}, 19\penalty0 (4):\penalty0 046006--046006, 2014.

\bibitem[Shaker et~al.(2017)Shaker, Monadjemi, Alirezaie, and Naghsh-Nilchi]{shaker2017dictionary}
Fariba Shaker, S~Amirhassan Monadjemi, Javad Alirezaie, and Ahmad~Reza Naghsh-Nilchi.
\newblock A dictionary learning approach for human sperm heads classification.
\newblock \emph{Computers in biology and medicine}, 91:\penalty0 181--190, 2017.

\bibitem[Shao et~al.(2023)Shao, Hu, Gao, Lei, Zhang, Meng, Xu, Huang, Li, Qiao, et~al.]{shao2023tiny}
Wenqi Shao, Yutao Hu, Peng Gao, Meng Lei, Kaipeng Zhang, Fanqing Meng, Peng Xu, Siyuan Huang, Hongsheng Li, Yu Qiao, et~al.
\newblock Tiny lvlm-ehub: Early multimodal experiments with bard.
\newblock \emph{arXiv preprint arXiv:2308.03729}, 2023.

\bibitem[Sharma et~al.(2018)Sharma, Ding, Goodman, and Soricut]{sharma2018conceptual}
Piyush Sharma, Nan Ding, Sebastian Goodman, and Radu Soricut.
\newblock Conceptual captions: A cleaned, hypernymed, image alt-text dataset for automatic image captioning.
\newblock In \emph{Proceedings of the 56th Annual Meeting of the Association for Computational Linguistics (Volume 1: Long Papers)}, pages 2556--2565, 2018.

\bibitem[Shiraishi et~al.(2000)Shiraishi, Katsuragawa, Ikezoe, Matsumoto, Kobayashi, Komatsu, Matsui, Fujita, Kodera, and Doi]{shiraishi2000development}
Junji Shiraishi, Shigehiko Katsuragawa, Junpei Ikezoe, Tsuneo Matsumoto, Takeshi Kobayashi, Ken-ichi Komatsu, Mitate Matsui, Hiroshi Fujita, Yoshie Kodera, and Kunio Doi.
\newblock Development of a digital image database for chest radiographs with and without a lung nodule: receiver operating characteristic analysis of radiologists' detection of pulmonary nodules.
\newblock \emph{American Journal of Roentgenology}, 174\penalty0 (1):\penalty0 71--74, 2000.

\bibitem[Soares et~al.(2020)Soares, Angelov, Biaso, Froes, and Abe]{soares2020sars}
Eduardo Soares, Plamen Angelov, Sarah Biaso, Michele~Higa Froes, and Daniel~Kanda Abe.
\newblock Sars-cov-2 ct-scan dataset: A large dataset of real patients ct scans for sars-cov-2 identification.
\newblock \emph{MedRxiv}, pages 2020--04, 2020.

\bibitem[Spanhol et~al.(2015)Spanhol, Oliveira, Petitjean, and Heutte]{spanhol2015dataset}
Fabio~A Spanhol, Luiz~S Oliveira, Caroline Petitjean, and Laurent Heutte.
\newblock A dataset for breast cancer histopathological image classification.
\newblock \emph{Ieee transactions on biomedical engineering}, 63\penalty0 (7):\penalty0 1455--1462, 2015.

\bibitem[Suckling(1994)]{suckling1994mammographic}
John Suckling.
\newblock The mammographic images analysis society digital mammogram database.
\newblock In \emph{Exerpta Medica. International Congress Series, 1994}, pages 375--378, 1994.

\bibitem[Tabik et~al.(2020)Tabik, G{\'o}mez-R{\'\i}os, Mart{\'\i}n-Rodr{\'\i}guez, Sevillano-Garc{\'\i}a, Rey-Area, Charte, Guirado, Su{\'a}rez, Luengo, Valero-Gonz{\'a}lez, et~al.]{tabik2020covidgr}
Siham Tabik, Anabel G{\'o}mez-R{\'\i}os, Jos{\'e}~Luis Mart{\'\i}n-Rodr{\'\i}guez, Iv{\'a}n Sevillano-Garc{\'\i}a, Manuel Rey-Area, David Charte, Emilio Guirado, Juan-Luis Su{\'a}rez, Juli{\'a}n Luengo, MA Valero-Gonz{\'a}lez, et~al.
\newblock Covidgr dataset and covid-sdnet methodology for predicting covid-19 based on chest x-ray images.
\newblock \emph{IEEE journal of biomedical and health informatics}, 24\penalty0 (12):\penalty0 3595--3605, 2020.

\bibitem[Touvron et~al.(2023)Touvron, Lavril, Izacard, Martinet, Lachaux, Lacroix, Rozi{\`e}re, Goyal, Hambro, Azhar, et~al.]{touvron2023llama}
Hugo Touvron, Thibaut Lavril, Gautier Izacard, Xavier Martinet, Marie-Anne Lachaux, Timoth{\'e}e Lacroix, Baptiste Rozi{\`e}re, Naman Goyal, Eric Hambro, Faisal Azhar, et~al.
\newblock Llama: Open and efficient foundation language models.
\newblock \emph{arXiv preprint arXiv:2302.13971}, 2023.

\bibitem[Tschandl et~al.(2018)Tschandl, Rosendahl, and Kittler]{tschandl2018ham10000}
Philipp Tschandl, Cliff Rosendahl, and Harald Kittler.
\newblock The ham10000 dataset, a large collection of multi-source dermatoscopic images of common pigmented skin lesions.
\newblock \emph{Scientific data}, 5\penalty0 (1):\penalty0 1--9, 2018.

\bibitem[Wang et~al.(2020)Wang, Lin, and Wong]{wang2020covid}
Linda Wang, Zhong~Qiu Lin, and Alexander Wong.
\newblock Covid-net: A tailored deep convolutional neural network design for detection of covid-19 cases from chest x-ray images.
\newblock \emph{Scientific reports}, 10\penalty0 (1):\penalty0 19549, 2020.

\bibitem[Wang et~al.(2022)Wang, Qin, Wang, Wang, Wang, Chen, Ouyang, Kuang, Dai, Mo, et~al.]{wang2022extreme}
Shuo Wang, Chen Qin, Chengyan Wang, Kang Wang, Haoran Wang, Chen Chen, Cheng Ouyang, Xutong Kuang, Chengliang Dai, Yuanhan Mo, et~al.
\newblock The extreme cardiac mri analysis challenge under respiratory motion (cmrxmotion).
\newblock \emph{arXiv preprint arXiv:2210.06385}, 2022.

\bibitem[Winzeck et~al.(2018)Winzeck, Hakim, McKinley, Pinto, Alves, Silva, Pisov, Krivov, Belyaev, Monteiro, et~al.]{winzeck2018isles}
Stefan Winzeck, Arsany Hakim, Richard McKinley, Jos{\'e}~AADSR Pinto, Victor Alves, Carlos Silva, Maxim Pisov, Egor Krivov, Mikhail Belyaev, Miguel Monteiro, et~al.
\newblock Isles 2016 and 2017-benchmarking ischemic stroke lesion outcome prediction based on multispectral mri.
\newblock \emph{Frontiers in neurology}, 9:\penalty0 679, 2018.

\bibitem[Wu et~al.(2023{\natexlab{a}})Wu, Lei, Zheng, Zhao, Lin, Zhang, Zhou, Zhao, Zhang, Wang, et~al.]{wu2023can}
Chaoyi Wu, Jiayu Lei, Qiaoyu Zheng, Weike Zhao, Weixiong Lin, Xiaoman Zhang, Xiao Zhou, Ziheng Zhao, Ya Zhang, Yanfeng Wang, et~al.
\newblock Can gpt-4v (ision) serve medical applications? case studies on gpt-4v for multimodal medical diagnosis.
\newblock \emph{arXiv preprint arXiv:2310.09909}, 2023{\natexlab{a}}.

\bibitem[Wu et~al.(2023{\natexlab{b}})Wu, Zhang, Zhang, Wang, and Xie]{wu2023towards}
Chaoyi Wu, Xiaoman Zhang, Ya Zhang, Yanfeng Wang, and Weidi Xie.
\newblock Towards generalist foundation model for radiology.
\newblock \emph{arXiv preprint arXiv:2308.02463}, 2023{\natexlab{b}}.

\bibitem[Xiong et~al.(2021)Xiong, Xia, Hu, Huang, Bian, Zheng, Vesal, Ravikumar, Maier, Yang, et~al.]{xiong2021global}
Zhaohan Xiong, Qing Xia, Zhiqiang Hu, Ning Huang, Cheng Bian, Yefeng Zheng, Sulaiman Vesal, Nishant Ravikumar, Andreas Maier, Xin Yang, et~al.
\newblock A global benchmark of algorithms for segmenting the left atrium from late gadolinium-enhanced cardiac magnetic resonance imaging.
\newblock \emph{Medical image analysis}, 67:\penalty0 101832, 2021.

\bibitem[Xu et~al.(2021)Xu, Zhu, Tang, Wang, Zhang, Li, Jiang, Shi, Liu, and Jin]{xu2021predicting}
Feng Xu, Chuang Zhu, Wenqi Tang, Ying Wang, Yu Zhang, Jie Li, Hongchuan Jiang, Zhongyue Shi, Jun Liu, and Mulan Jin.
\newblock Predicting axillary lymph node metastasis in early breast cancer using deep learning on primary tumor biopsy slides.
\newblock \emph{Frontiers in oncology}, 11:\penalty0 759007, 2021.

\bibitem[Xu et~al.(2023{\natexlab{a}})Xu, Shao, Zhang, Gao, Liu, Lei, Meng, Huang, Qiao, and Luo]{xu2023lvlm}
Peng Xu, Wenqi Shao, Kaipeng Zhang, Peng Gao, Shuo Liu, Meng Lei, Fanqing Meng, Siyuan Huang, Yu Qiao, and Ping Luo.
\newblock Lvlm-ehub: A comprehensive evaluation benchmark for large vision-language models.
\newblock \emph{arXiv preprint arXiv:2306.09265}, 2023{\natexlab{a}}.

\bibitem[Xu et~al.(2023{\natexlab{b}})Xu, Zhang, Xie, Zhao, Guo, Wong, Li, and Zhao]{xu2023drivegpt4}
Zhenhua Xu, Yujia Zhang, Enze Xie, Zhen Zhao, Yong Guo, Kenneth~KY Wong, Zhenguo Li, and Hengshuang Zhao.
\newblock Drivegpt4: Interpretable end-to-end autonomous driving via large language model.
\newblock \emph{arXiv preprint arXiv:2310.01412}, 2023{\natexlab{b}}.

\bibitem[Xuehai et~al.(2020)Xuehai, Yichen, Luntian, Eric, and Pengtao]{xuehai2020pathvqa}
He Xuehai, Zhang Yichen, Mou Luntian, Xing Eric, and Xie Pengtao.
\newblock Pathvqa: 30000+ questions for medical visual question answering.
\newblock \emph{arXiv preprint arXiv:2003.10286}, 2020.

\bibitem[Ye et~al.(2023)Ye, Xu, Xu, Ye, Yan, Zhou, Wang, Hu, Shi, Shi, et~al.]{ye2023mplug}
Qinghao Ye, Haiyang Xu, Guohai Xu, Jiabo Ye, Ming Yan, Yiyang Zhou, Junyang Wang, Anwen Hu, Pengcheng Shi, Yaya Shi, et~al.
\newblock mplug-owl: Modularization empowers large language models with multimodality.
\newblock \emph{arXiv preprint arXiv:2304.14178}, 2023.

\bibitem[Zawacki et~al.(2019)Zawacki, Wu, Shih, Elliott, Fomitchev, Hussain, Lakhani, Culliton, and Bao]{zawacki2019siim}
Anna Zawacki, Carol Wu, George Shih, Julia Elliott, Mikhail Fomitchev, Mohannad Hussain, Paras Lakhani, Phil Culliton, and Shunxing Bao.
\newblock Siim-acr pneumothorax segmentation.
\newblock \url{https://kaggle.com/competitions/siim-acr-pneumothorax-segmentation}, 2019.

\bibitem[Zhang et~al.(2023{\natexlab{a}})Zhang, Fei, Yao, Ji, Li, Liu, and Chua]{zhang2023transfer}
Ao Zhang, Hao Fei, Yuan Yao, Wei Ji, Li Li, Zhiyuan Liu, and Tat-Seng Chua.
\newblock Transfer visual prompt generator across llms.
\newblock \emph{arXiv preprint arXiv:2305.01278}, 2023{\natexlab{a}}.

\bibitem[Zhang et~al.(2023{\natexlab{b}})Zhang, Xu, Usuyama, Bagga, Tinn, Preston, Rao, Wei, Valluri, Wong, et~al.]{zhang2023large}
Sheng Zhang, Yanbo Xu, Naoto Usuyama, Jaspreet Bagga, Robert Tinn, Sam Preston, Rajesh Rao, Mu Wei, Naveen Valluri, Cliff Wong, et~al.
\newblock Large-scale domain-specific pretraining for biomedical vision-language processing.
\newblock \emph{arXiv preprint arXiv:2303.00915}, 2023{\natexlab{b}}.

\bibitem[Zhang et~al.(2023{\natexlab{c}})Zhang, Wu, Zhao, Lin, Zhang, Wang, and Xie]{zhang2023pmc}
Xiaoman Zhang, Chaoyi Wu, Ziheng Zhao, Weixiong Lin, Ya Zhang, Yanfeng Wang, and Weidi Xie.
\newblock Pmc-vqa: Visual instruction tuning for medical visual question answering.
\newblock \emph{arXiv preprint arXiv:2305.10415}, 2023{\natexlab{c}}.

\bibitem[Zhu et~al.(2023)Zhu, Chen, Shen, Li, and Elhoseiny]{zhu2023minigpt}
Deyao Zhu, Jun Chen, Xiaoqian Shen, Xiang Li, and Mohamed Elhoseiny.
\newblock Minigpt-4: Enhancing vision-language understanding with advanced large language models.
\newblock \emph{arXiv preprint arXiv:2304.10592}, 2023.

\end{thebibliography}
}


\end{document}